\documentclass[a4paper,11pt]{article}
\usepackage[utf8]{inputenc}
\usepackage[left=3cm,
            right=2.5cm,
            top=2.5cm,
            bottom=3cm
            ]{geometry} 

\usepackage{tcolorbox}
\usepackage{amsmath,amsfonts,amssymb,array,graphicx,color,dsfont,mathtools,mathrsfs}
\usepackage{physics}
\usepackage{bbm}
\usepackage{hyperref}
\hypersetup{colorlinks,bookmarksopen,bookmarksnumbered,
 citecolor=magenta,
 linkcolor=blue}
 
\usepackage{blkarray}
\usepackage{comment}
\usepackage{authblk}
\usepackage{appendix}

\numberwithin{equation}{section}

\usepackage[style=phys,hyperref=true,articletitle=true,biblabel=brackets,chaptertitle=false]{biblatex}
\usepackage{braket}  
\def \be {\begin{equation}} 
\def \ee {\end{equation}} 
\def \l {\left(} 
\def \r {\right)} 
\def \la {\langle} 
\def \ra {\rangle}  

\date{}

\title{
Entanglement content of kink excitations
}
\author{Luca Capizzi$^{1}$, Michele Mazzoni$^{2}$}

\begin{document}

\maketitle

$^1$Universit\'e Paris-Saclay, CNRS, LPTMS, 91405, Orsay, France.\\
$^{2}$Department of Mathematics, City, University of London, 10 Northampton Square EC1V 0HB, UK.\\

\begin{abstract}

Quantum one-dimensional systems in their ordered phase admit kinks as elementary excitations above their symmetry-broken vacua. While the scattering properties of the kinks resemble those of quasiparticles, they have distinct locality features that are manifest in their entanglement content. In this work, we study the entanglement entropy of kink excitations. We first present detailed calculations for specific states of a spin-1/2 chain to highlight the salient features of these excitations. Second, we provide a field-theoretic framework based on the algebraic relations between the twist fields and the semilocal fields associated with the excitations, and we compute the R\'enyi entropies in this framework. We obtain universal predictions for the entropy difference between the excited states with a finite number of kinks and the symmetry-broken ground states, which do not depend on the microscopic details of the model in the limit of large regions. Finally, we discuss some consequences of the Kramers-Wannier duality, which relates the ordered and disordered phases of the Ising model, and we explain why, counterintuitively, no explicit relations between those phases are found at the level of entanglement.
\end{abstract}

\tableofcontents 

\section{Introduction}

In the last decades, several studies regarding the entanglement of quantum systems shed light on the fundamental properties of quantum correlations in the ground state of many-body systems \cite{afov-08}. In particular, many exact results have been provided in the context of critical ground states via Conformal Field Theory (CFT) techniques \cite{cw-93,hlw-94,cc-09}. Moreover, analytical investigations of gapped systems close to criticality have been given in a series of works by means of form-factor bootstrap for integrable systems \cite{ccd-08,cd-08,cd-09,bc-16,CastroAlvaredo-17}. Despite the abundance of works for quantum systems with a unique vacuum, for instance, many-body models in their disordered phase, only a few results are present for the symmetry-broken ground states, especially in the context of Quantum Field Theory (QFT). Considering the importance of spontaneous symmetry breaking in modern physics, this fact is quite surprising.

 Since there are Kramers-Wannier dualities connecting ordered and disordered phases \cite{Savit-80}, one could erroneously argue that there should be no discernible distinction in the entanglement content between these phases, and consequently identical predictions should arise at corresponding dual points. However, this is not the case, as previous exact lattice results for the Ising chain pointed out \cite{ccp-10}. While the presence of these discrepancies is known (see also Ref. \cite{dh-23}), some fundamental issues regarding their origin are not completely understood: is there a relation between the ground state entanglement of ordered and disordered phases? Is there a quantity dual to the entanglement entropy under Kramers-Wannier?

In a previous work \cite{cm-23}, we have shown how the universal corrections to the area law found in Ref. \cite{Doyon-09} for a large class of theories with a single vacuum do not apply to the Ising Field Theory in its ordered phase (where spontaneous symmetry breaking arises and two vacua are present); that was traced back to the topological nature of the kink excitations and the corresponding selection rules for the form factors. A natural question, that is the focus of this work, is to understand whether similar discrepancies can be found for the low-lying excited states as well. Namely, in the series of works \cite{cdds-18a,cdds-18b,cdds-19a,cdds-19b} the excess of entropy for the particle excitations in the disordered phase of a 1+1D gapped many-body system was proved to be universal. Those predictions boil down to general formulae with a simple semiclassical interpretation, and we ask whether similar results can be found for the ordered phase. We anticipate that we will find universal results different from those of the disordered phase; discrepancies are manifest, for example, in the R\'enyi entropies of tripartite geometries in the presence of a single kink.

It is worth mentioning that the importance of the intrinsic non-locality of the kinks has been remarked for the quench dynamics in quantum spin chains in Refs. \cite{em-18,em-20,ge-21,ds-22,ds-23,kkhpk-23,kt-24}. Moreover, topological frustration, as for antiferromagnetic chains with an odd number of sites, forces the ground state to be a one-kink state, with many counterintuitive consequences highlighted in Refs. \cite{mgf-20,mgkf-20,grf-19,tofgf-23}. Therefore, a deeper understanding of the entanglement content of kinks is imperative to correctly interpret a plethora of phenomena arising in one-dimensional quantum systems.

We organize the manuscript as follows. In Section \ref{sec:lattice_qubit} we first introduce the problem and obtain some explicit results on the lattice, and then, in the spirit of \cite{cdds-18a}, we provide a qubit picture that is sufficient to infer the correct universal contribution to the entanglement entropy arising from the kinks. Section \ref{Sec:field-theoretic-approach}, which is the core of our work, contains a field-theoretic explanation of the mechanism. In particular, we make use of the replica trick to compute the R\'enyi entropies via the so-called twist fields, and we trace back the origin of the novel entanglement properties to the algebraic relations between the twist fields and the disorder lines associated with the kinks. Finally, in Section \ref{Sec:Kramers-Wannier and entanglement of algebras} we analyze the Kramers-Wannier duality and we explain the faith of the density matrix associated with local regions under the duality within a $C^*$-algebraic approach; this allows us to understand in detail why the entropy is not self-dual and we give an explicit construction of its dual. Conclusions are given in Section \ref{Sec:conclusion}.

\section{Predictions from spin chains and the qubit picture}
\label{sec:lattice_qubit}

In this section, we compute the R\'enyi entropies of states of a 1D spin chain containing one or two kink excitations and we show that, in the case of a spatial tripartition, our predictions differ from those obtained for localised quasiparticle excitations. For the latter, the excess of R\'enyi and entanglement entropy with respect to the ground state is well known in the case of CFTs \cite{alcaraz2011entanglement, berganza2012entanglement}, massive integrable QFTs \cite{cdds-18a,cdds-18b,cdds-19a,cdds-19b} and spin chains \cite{jafarizadeh2019bipartite, zhang2021universal,zhang2021corrections,zhang2022entanglement,molter2014bound}. In particular, in a state of a massive QFT containing a single localised excitation, the excess of entanglement entropy of a region $A$ with fixed relative volume $r= V_A / V$ in the large volume limit is given by 
\be
\label{eq:local_quasiparticle_entropy}
\Delta S = -r \log r - (1-r)\log(1-r).
\ee
This result relies on the locality of the excitation, and it also applies whenever $A$ is made of disjoint connected regions. On the other hand, if we consider the topological excitations of a spin chain, such as the kinks in the ordered phase of the Ising model, the situation is different. Indeed, these excitations separate regions of different magnetisations, and thus by measuring the local magnetisation at a given point one could tell, in principle, whether the domain wall is found to the right or to the left of that point. The same is not possible for (local) quasiparticles, and physically one expects that this is the main feature responsible for the difference between kinks and particles at the level of entanglement content: in particular, we will show that \eqref{eq:local_quasiparticle_entropy} does not hold for kinks if $A$ consists of multiple disjoint regions.

\begin{figure}[t]
    \centering
    \includegraphics[width=0.7\linewidth]{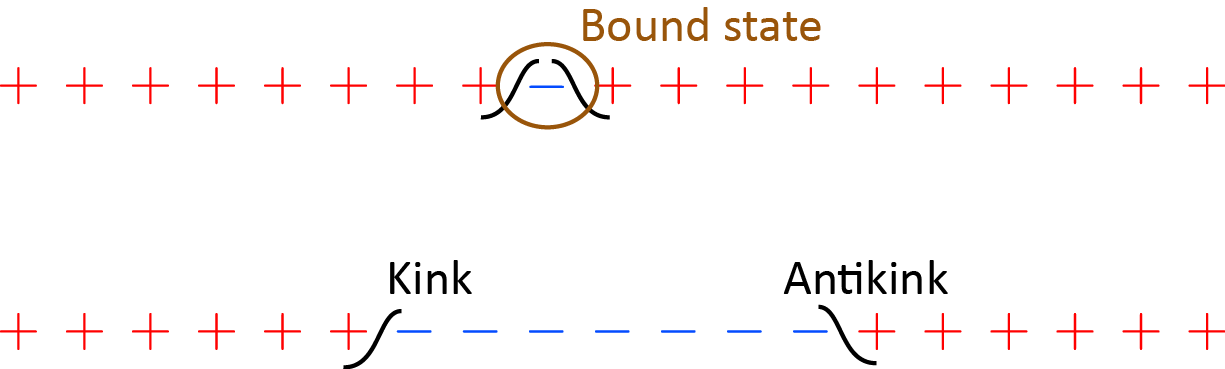}
    \caption{Sketch of a two-kink state. Top: the two kinks are bound together and they behave as a single particle. Bottom: A kink and an antikink are deconfined along the system, and they interpolate between the two vacua.}
    \label{fig:kinks1}
\end{figure}

The situation becomes more complex in the presence of two or more kinks, leading to a richer phenomenology. In particular, if two kinks are sufficiently close they behave as a single collective particle (bound state): measuring the magnetisation at a distant point does not give information on the position of the bound state. Conversely, if the kinks are well separated, a measurement at a given point can reveal whether the two kinks belong to the opposite sides of the system. We depict the two scenarios in Figure \ref{fig:kinks1}.

In this section, we study the low-lying spectrum of a spin chain deep in its order phase (e.g. the quantum Ising chain at small transverse field), where the two vacua become product states with opposite magnetisation and the kinks are plane-wave superpositions of domain wall configurations: this limiting regime, where the correlation functions of the vacua shrink to zero, is paradigmatic in its simplicity and it allows to elucidate the difference between quasiparticles and kinks. For states with one or two kinks, we are able to perform exact calculations on the lattice. However, since the task becomes challenging as the number of kinks increases, we also propose an alternative picture based on multi-qubit states; we adapt the qubit picture of Refs. \cite{cdds-18a,cdds-18b,cdds-19a,cdds-19b}, initially introduced to describe the entanglement entropy and negativity of localised excitations, to deal with kinks.

\subsection{R\'enyi entropies of some one- and two-kink states in a spin-$\frac{1}{2}$ chain}\label{sec:spin_entropy}

\subsubsection{Structure of the Hilbert space}
\label{sec:Structure of the Hilbert space}

We consider a spin-$\frac{1}{2}$ chain of length $L$, whose Hilbert space is
\be
\label{eq:spin_one_half_Hilbert_space}
\mathcal{H} = \text{Span}\{ \ket{s_1,\dots,s_L} , \quad s_j = \pm\} \simeq \l\mathbb{C}^2\r^{\otimes L}.
\ee
We regard the two states $\ket{+\dots +}$ and $\ket{-\dots -}$ as vacua, related by a global $\mathbb{Z}_2$ symmetry, and we parametrise the other configurations via domain walls at different lattice sites as explained below. 

First, according to the sign of the spin on the left boundary, we identify two sectors (of dimension $2^{L-1}$) defined as
\be
\mathcal{H}^{(\pm)} \equiv \text{Span}\{ \ket{\pm,s_2,\dots,s_L}, \quad s_j = \pm \},
\ee
and such that $\mathcal{H} = \mathcal{H}^{(+)} \oplus \mathcal{H}^{(-)}$. Additionally, we split $\mathcal{H}^{(\pm)}$ into sectors with a fixed number of ($N$) domain walls as
\be
\mathcal{H}^{(\pm)} = \bigoplus^{L-1}_{N=0} \mathcal{H}^{(\pm)}_N.
\ee
Specifically, $\mathcal{H}^{(\pm)}_N$ is generated by the configurations starting from $\pm$ with $N$ changes of sign (domain walls) between neighboring sites.

In the following, without loss of generality, we will focus on the sector $\mathcal{H}^{(+)}$, with the first spin being up. For $N=0$, $\mathcal{H}^{(\pm)}_{0}$ is generated by $\ket{+,\dots,+}$, meaning that no domain walls are present. For $N=1$, $\mathcal{H}^{(+)}_1$ admits the following basis
\be
\ket{K_{+-}(j)} \equiv \ket{+,+,\dots,+,-,\dots,-}, \quad j=1,\dots,L-1,
\ee
with a domain wall between the positions $j$ and $j+1$. Similarly, for $N=2$, $\mathcal{H}^{(+)}_2$ is spanned by the states
\be
\ket{K_{+-}(j_1)K_{-+}(j_2)} \equiv \ket{+,+\dots,+,-,\dots,-,+,\dots,+}, \quad  1 \le j_1 <j_2 \le L-1,
\ee
with the first domain wall between sites $j_1$, $j_1 +1$, and the second between sites $j_2$ and $j_2 +1$. More generally, a basis for $\mathcal{H}^{(+)}_N$ is parametrised by sequences of $N$ ordered positions $(j_1,\dots,j_N)$, belonging to $\{1,\dots,L-1\}$, and therefore the dimension of this sector is
\be
\text{dim}\mathcal{H}^{(+)}_N = \binom{L-1}{N};
\ee
this is consistent with $\sum^{L-1}_{N=0}\text{dim}\mathcal{H}^{(+)}_N = \text{dim}\mathcal{H}^{(+)}$, since $\sum_{N=0}^{L-1}\binom{L-1}{N}=2^{L-1}$.


\subsubsection{One-kink state} 

In this section, we analyse a state given by a coherent superposition of configurations with a single domain wall:
\be
\ket{\mathcal{K}_{+-}(p)} = \frac{1}{\sqrt{L-1}} \sum^{L-1}_{j=1}e^{i p j} \ket{K_{+-}(j)}.
\ee
This state represents a kink with a given momentum $p$ that is completely delocalised in space. 

We are interested in the computation of the R\'enyi entropy of extended regions, and we first analyse the simplest case of a bipartition $A\cup \bar{A}$, with
\be\label{eq:bipartite_A}
A = \{1,\dots,\ell\},\quad \bar{A} = \{\ell +1,\dots L\}.
\ee
We can compute the reduced density matrix (RDM) by tracing out the degrees of freedom in $\bar{A}$:
\be
\rho^{(1)}_A = \text{Tr}_{\bar{A}}\l \ket{\mathcal{K}_{+-}(p)} \bra{\mathcal{K}_{+-}(p)}\r \equiv  \underset{s_{\ell + 1},\dots,s_L = \pm}{\sum} \prescript{}{\bar{A}}{\braket{s_{\ell +1},\dots,s_L | \mathcal{K}_{+-}(p)}}\braket{\mathcal{K}_{+-}(p)| s_{\ell +1},\dots,s_L}_{\bar{A}},
\ee
obtaining
\begin{align}
\label{eq:one_kink_bipartite_RDM}
\rho^{(1)}_A &= \frac{1}{L-1}\sum_{j,j'=1}^{\ell-1} e^{i p (j-j')}\ket{K_{+-}(j)}_A \prescript{}{A}{\bra{K_{+-}(j')}} + \l1-\frac{\ell}{L-1}\r \ket{+,\dots,+}_A \prescript{}{A}{\bra{+,\dots,+}} \nonumber \\
&+\frac{1}{L-1}\sum_{j=1}^{\ell-1}\l e^{i p (j-\ell)}\ket{K_{+-}(j)}_A \prescript{}{A}{\bra{+,\dots,+}} \,+\, \text{h.c.}\r,
\end{align}
with $\ket{K_{+-}(j)}_A$ the state of $A$ with a kink at position $j$. We identify three types of contributions in Eq. \eqref{eq:one_kink_bipartite_RDM}: a term with one kink in $A$, a term with no kinks in $A$, and a mixed term that couples the two sectors; in particular, one can show that the latter contribution is due to the configurations with a domain wall lying between $\ell$ and $\ell+1$, that is the edge of the region $A$. While the exact diagonalisation of $\rho^{(1)}_A$ at finite size is quite cumbersome, a simplification occurs in the limit
\be
\label{eq:large_volume_limit_bipartite}
\ell \to \infty, \quad L \to \infty, \quad  r\equiv \frac{\ell}{L} \quad \text{fixed}.
\ee
In particular, the first two terms of \eqref{eq:one_kink_bipartite_RDM} can be diagonalised simultaneously and the associated non-vanishing eigenvalues are $r$ and $1-r$ respectively: the third term can be treated in perturbation theory, and in the limit \eqref{eq:large_volume_limit_bipartite} one can easily show that the associated contribution to the spectrum goes to zero. Therefore, in this limit, the $n$th R\'enyi entropy of the one-kink state is
\be\label{entropy_1K_quasiparticle_prediction}
S_n = \frac{1}{1-n}\log \Tr\big[(\rho_A^{(1)})^n\big] \simeq \frac{1}{1-n}\log(r^n + (1-r)^n).
\ee
We observe that this result coincides with the predictions of Ref. \cite{cdds-18a} for quasiparticles; in particular, the entanglement entropy, obtained in the limit $n\rightarrow 1$ of Eq. \eqref{entropy_1K_quasiparticle_prediction}, directly yields \eqref{eq:local_quasiparticle_entropy} since the entropy of the vacuum state $\ket{+,\dots,+}$ vanishes. This means that in the case of the geometry $\eqref{eq:bipartite_A}$ there is no distinction between kinks and quasiparticle at the level of entanglement entropy (at least, in the large volume limit).

We now turn to a tripartite geometry, and we consider the entanglement between a single interval and the rest. Namely, we choose $A = A_1\cup A_3,\, \bar{A}=A_2$ with
\be
\label{eq:tripartite_geometry}
A_1 = \{1,\dots,\ell_1\}, \quad A_2 = \{\ell_1 +1,\dots,\ell_1 + \ell_2\}, \quad A_3 = \{\ell_1 + \ell_2 +1,\dots,L\};
\ee
therefore, after defining $\ell_3 \equiv L-\ell_1-\ell_2$, the length of the interval $A_i$ is $\ell_i$.
The evaluation of the reduced density matrix $\rho^{(1)}_{A_1\cup A_3}$ is analogous to the bipartite case discussed above. Thus, after some algebra, we obtain the following expression
\begin{align}
\label{eq:1k_tripartite_RDM}
&\rho^{(1)}_{A_1 \cup A_3} = \frac{1}{L-1}\sum_{j,j'=1}^{\ell_1-1} e^{ip (j-j')}\ket{K_{+-}(j)}_{A_1} \prescript{}{A_1}{\bra{K_{+-}(j')}} \otimes \ket{-\dots,-}_{A_3} \prescript{}{A_3}{\bra{-,\dots,-}} \nonumber \\ &+ \frac{1}{L-1}\sum_{j,j'=\ell_1+\ell_2 +1}^{L-1} e^{ip(j-j')}\ket{+,\dots,+}_{A_1} \prescript{}{A_1}{\bra{+,\dots,+}} \otimes \ket{K_{+-}(j)}_{A_3} \prescript{}{A_3}{\bra{K_{+-}(j')}} \nonumber \\&+ \frac{\ell_2 +1}{L-1} \ket{+,\dots,+}_{A_1} \prescript{}{A_1}{\bra{+,\dots,+}} \otimes \ket{-\dots,-}_{A_3} \prescript{}{A_3}{\bra{-,\dots,-}}
\nonumber \\&+ \frac{1}{L-1} \sum_{j=1}^{\ell_1 -1}\l e^{i p (j-\ell_1)}\ket{K_{+-}(j)}_{A_1} \prescript{}{A_1}{\bra{+,\dots,+}} \otimes \ket{-\dots,-}_{A_3} \prescript{}{A_3}{\bra{-,\dots,-}}\,+\, \text{h.c.}\r
\nonumber \\&+ \frac{1}{L-1} \sum_{j=\ell_1 + \ell_2 + 1}^{L-1}\l e^{i p (j-\ell_1-\ell_2)}\ket{+,\dots,+}_{A_1} \prescript{}{A_1}{\bra{+,\dots,+}} \otimes \ket{-\dots,-}_{A_3} \prescript{}{A_3}{\bra{K_{+-}(j)}}\,+\, \text{h.c.}\r.
\end{align}
As expected, the number of kinks belonging to $A_1 \cup A_3$ can be either $0$ or $1$, and we identify the first three terms in \eqref{eq:1k_tripartite_RDM} as those associated with a fixed number of kinks in the regions $A_1$ and $A_3$; specifically, after decomposing the Hilbert space $\mathcal{H}_{A_1}\otimes \mathcal{H}_{A_3}$ in sectors, we find
\begin{itemize}
    \item One $(\ell_1 -1) \times (\ell_1-1)$ block acting on $ \mathcal{H}_{A_1,1}^{(+)} \otimes \mathcal{H}_{A_3,0}^{(-)}$  (one kink in $A_1$, no kinks in $A_3$),
    \item One $(\ell_3 -1) \times (\ell_3-1)$ block acting on $\mathcal{H}_{A_1,0}^{(+)} \otimes \mathcal{H}_{A_3,1}^{(+)}$  (no kinks in $A_1$, one kink in $A_3$),
    \item One $1 \times 1$ block on acting $\mathcal{H}_{A_1,0}^{(+)} \otimes \mathcal{H}_{A_3,0}^{(-)}$ (no kinks in $A$).
\end{itemize}
On the other hand, the last two terms of \eqref{eq:1k_tripartite_RDM} mix the sectors. However, in the limit $L\rightarrow \infty$ with
\be
\label{eq:large_volume_limit_tripartite}
r_i \equiv \frac{\ell_i}{L} \text{ fixed}, \quad i=1,2,3;
\ee
the eigenvalues of the first three blocks converge to $r_1$, $r_2$, $r_3$ respectively, and the contribution coming from the two mixing terms goes to zero. In conclusion, we compute the R\'enyi entropy as
\be
\label{eq:entropy_1K_tripartite_lattice}
S_n = \frac{1}{1-n}\log \Tr\big[(\rho_{A_1\cup A_3}^{(1)})^n\big] \simeq \frac{1}{1-n}\log(r_1^n + r_2^n + r_3^n),
\ee
which coincides with \eqref{entropy_1K_quasiparticle_prediction} for $r_3=0$. As anticipated, this result differs from the one of quasiparticles in a tripartite geometry \eqref{eq:local_quasiparticle_entropy}, where no distinction between connected and disconnected regions is found. The difference between the two results is a hallmark of the non-locality of kink excitations, and it is physically better understood when one takes the limit $r_2 \to 0$: we shall come back to this point at the end of Section \ref{Sec:field-theoretic-approach}.

\subsubsection{Two-kink states}\label{sec:two-kink}

 Interesting features related to the non-locality of kink excitations emerge in the presence of two kinks for the tripartite geometry in Eq. \eqref{eq:tripartite_geometry}. As a paradigmatic example, we focus on the case where the two kinks have equal momenta, $p=p'=0$, and we consider the state
 \be\label{two_kinks_delocalised}
\ket{\mathcal{K}_{+-}(0)\mathcal{K}_{-+}(0)}=\sqrt{\frac{2}{(L-1)(L-2)}}\sum_{1\le j < j' \le L-1}\ket{K_{+-}(j)K_{-+}(j')},
\ee 
where the prefactor ensures the correct normalisation. The computation of its reduced density matrix, denoted by $\rho^{(2)}_{A_1 \cup A_3}$, does not pose additional difficulties, but it is lengthy and requires some care. We leave the details of the computation and the explicit expression of $\rho^{(2)}_{A_1 \cup A_3}$ to Appendix \ref{app:RDM of the lattice two-kink state}; here, we discuss its salient features.

First, since two kinks are present in the entire system, only zero, one, or two kinks can be present in $A_1 \cup A_3$. Second, in analogy to the previous calculation, only the contributions coming from the blocks with a fixed number of kinks in $A$ matter in the limit \eqref{eq:large_volume_limit_tripartite}; we report them below, together with their non-zero eigenvalues:
\begin{itemize}
    \item One $1 \times 1$ block acting on $\mathcal{H}_{A_1,0}^{(+)} \otimes \mathcal{H}_{A_3,0}^{(+)}$ (no kinks in $A$). It yields an eigenvalue $r_2^2$. 
    \item One $(\ell_1 -1) \times (\ell_1-1)$ block acting on $ \mathcal{H}_{A_1,1}^{(+)} \otimes \mathcal{H}_{A_3,0}^{(+)}$  (one kink in $A_1$, one in $\bar{A}$). It yields an eigenvalue $2r_1 r_2$.
    \item One $(\ell_3 -1) \times (\ell_3-1)$ block acting on $ \mathcal{H}_{A_1,0}^{(+)} \otimes \mathcal{H}_{A_3,1}^{(-)}$  (one kink in $\bar{A}$, one in $A_3$). It yields an eigenvalue $2r_3 r_2$.
    \item One $(\ell_1 -1)(\ell_3-1) \times (\ell_1 -1)(\ell_3-1)$ block acting on $ \mathcal{H}_{A_1,1}^{(+)} \otimes \mathcal{H}_{A_3,1}^{(-)}$  (one kink in $A_1$ and one in $A_3$). It yields an eigenvalue $2r_1 r_3$.
    \item One $[(\ell_1 -1)(\ell_1-2)/2 + (\ell_3 -1)(\ell_3-2)/2] \times[(\ell_1 -1)(\ell_1-2)/2 + (\ell_3 -1)(\ell_3-2)/2] $ block acting on\footnote{Here, the tensor product $\otimes$ is performed first, followed by the direct sum $\oplus$.} $ \mathcal{H}_{A_1,2}^{(+)} \otimes \mathcal{H}_{A_3,0}^{(+)} \oplus \mathcal{H}_{A_1,0}^{(+)} \otimes \mathcal{H}_{A_3,2}^{(+)}$ (two kinks in $A$). It yields an eigenvalue $r_1^2 + r_3^2$.
\end{itemize}

It is worth noting that the last block gives rise to a mixing of two sectors, corresponding to the two kinks being in either $A_1$ or $A_3$, respectively. From the point of view of entanglement, this contribution can be interpreted as the two kinks behaving like a single bound state which has a probability $r_1^2 + r_3^2$ of being found in $A_1 \cup A_3$. Similarly, the first block corresponds to a single bound state belonging to $A_2$ with probability $r^2_2$. Conversely, in all the other cases the kinks belong to distinct regions, and their genuine non-locality manifests explicitly in their entanglement contributions.

We conclude by reporting the value of the R\'enyi entropy for the tripartition in the large volume limit:
\be
S_n \simeq \frac{1}{1-n}\log(r_2^{2n} + (r_1^2 + r_3^2)^n + 2^n(r_1^n r_2^n + r_1^n r_3^n+r_2^n r_3^n)).
\ee
This expression is different from that of the two quasiparticle excitations in Ref. \cite{cdds-19a}, which depends only on the sum $r_1 + r_3$ and not on $r_1$ and $r_3$ separately.

\subsection{The qubit picture}
In this section we further explore the difference between multi-kink and multi-particle states at the level of entanglement by adopting the qubit formalism, first introduced in \cite{cdds-18a,cdds-18b,cdds-19a,cdds-19b}, and later extended to deal with internal symmetries \cite{capizzi2022symmetry, cdmsc-22,capizzi2023symmetry}. The idea is to recast the states considered in Section \eqref{sec:spin_entropy}, into simpler states by retaining the same entanglement content (at least, in the large volume limit). Before showing how to generalise the formalism to kinks, we first review the qubit description in the case of quasiparticle excitations.

\subsubsection{One- and two-particle states, revisited}
Let us consider the same tripartite geometry of Section \eqref{eq:tripartite_geometry}, and we focus on a single quasiparticle at a given momentum (completely delocalised in space). We assign $1$ to the region $A_i$ whenever the particle belongs to it, and $0$ otherwise. The qubit representation of the state above is
\be
\ket{\Psi^{(1)}} = \sqrt{r_1} \ket{100} + \sqrt{r_2}\ket{010} + \sqrt{r_3} \ket{001},
\ee
so that $r_i$ is the probability that the particle belongs to $A_i$.
Its RDM is
\be
\rho_A^{(1)} = \Tr_{A_2} \ket{\Psi^{(1)}}\bra{\Psi^{(1)}} =
\l
\begin{array}{c|ccc}
&\ket{00} & \ket{10} & \ket{01} \\
\hline
\bra{00} & r_2 & 0 & 0 \\
\bra{10} & 0 & r_1 & \sqrt{r_1 r_3}\\
\bra{01} & 0 & \sqrt{r_1 r_3} & r_3 \\
\end{array}
\r.
\ee
After introducing
\be
\ket{\Phi_0} = \ket{00}, \quad \ket{\Phi_1} = \frac{\sqrt{r_1}\ket{10}+\sqrt{r_3}\ket{01}}{\sqrt{r_1 + r_3}},
\ee
we express the RDM in terms of its spectral projectors as
\be
\rho_A^{(1)} = (1-r)\ket{\Phi_0}\bra{\Phi_0} + r\ket{\Phi_1}\bra{\Phi_1},
\ee
with $r=r_1+r_3$. In particular, the entanglement content, encoded in the non-zero eigenvalues of $\rho^{(1)}_A$ ($r$ and $1-r$) depends only on the probability $r$ that the particle belongs to $A=A_1\cup A_3$.

The situation is slightly more complex in the presence of two indistinguishable particles. In that case, we assign the value $2$ to the region $A_i$ if both particles belong to $A_i$ and we express the qubit-state as
\be
\ket{\Psi^{(2)}} = r_1 \ket{200} + r_2\ket{020} + r_3 \ket{002} + \sqrt{2r_1 r_2}\ket{110} + \sqrt{2r_1 r_3}\ket{101}+\sqrt{2r_2 r_3}\ket{011},
\ee
whose corresponding RDM is
\begin{align}
\label{2P_qb_RDM}
\rho_A^{(2)} &= \Tr_{A_2} \ket{\Psi^{(2)}}\bra{\Psi^{(2)}} \nonumber \\&=
\l
\begin{array}{c|cccccc}
&\ket{00} & \ket{10} & \ket{01} &\ket{20} & \ket{02} & \ket{11} \\
\hline
\bra{00} & r_2^2 & 0 & 0 & 0 & 0 & 0 \\
\bra{10} & 0 & 2 r_2 r_1 & 2 r_2\sqrt{r_1 r_3} & 0 & 0 & 0\\
\bra{01} & 0 & 2 r_2 \sqrt{r_1 r_3} & 2 r_2 r_3 & 0 & 0 & 0\\
\bra{20} & 0 & 0 & 0 & r_1^2 & r_1 r_3 & r_1 \sqrt{2 r_1 r_3} \\
\bra{02} & 0 & 0 & 0 & r_1 r_3 & r_3^2 & r_3 \sqrt{2 r_1 r_3} \\
\bra{11} & 0 & 0 & 0 & r_1 \sqrt{2 r_1 r_3} & r_3 \sqrt{2 r_1 r_3}& 2 r_1 r_3 \\
\end{array}
\r.
\end{align}
Its non-zero eigenvalues are $(1-r)^2,\,2r(1-r),\,r^2$ and, again, they only depend on $r_1,r_3$ though their sum $r$.
An important message is that mixing is always present between the states with a given number of particles in $A_1\cup A_3$: as we will see, this is not the case for the kinks.

\subsubsection{R\'enyi entropies of multi-kink states in a tripartite geometry}

We now adopt a qubit-like formalism to describe the states containing multiple kinks/antikinks. The key idea consists of fixing the number of kinks in a given region and then forgetting about all the other details related to the spatial distributions of spins in that region. To this aim, we introduce the states $\ket{\pm}$ to represent the two vacua, and $\ket{1^\pm}$ as a kink (or antikink) interpolating between $\ket{\pm}$ and $\ket{\mp}$. The state of a single kink delocalised in space is
\be
\ket{\Psi^{(1)}} = \sqrt{r_1} \ket{1^+, -, -} +\sqrt{r_2} \ket{+, 1^+, -}+ \sqrt{r_3} \ket{+, +, 1^+},
\ee
and its RDM for the region $A=A_1\cup A_3$ is
\begin{align}
    \rho_A^{(1)} = r_1 \ket{1^+,-}\bra{1^+,-} + r_2 \ket{+,-}\bra{+,-} + r_3 \ket{+,1^+}\bra{+,1^+}.
\end{align}
Crucially, this result reproduces the same non-zero spectrum of \eqref{eq:tripartite_geometry} in the large-volume limit and it shows the absence of mixing between the states with a single kink in $A_1$ or $A_3$ respectively (in contrast with the quasiparticle state in Eq. \eqref{2P_qb_RDM}). 

We now focus on a two-kink state. First, we introduce the state $\ket{2^{\pm}}$ to represent the presence of a kink-antikink pair interpolating between $\ket{\pm}$ and $\ket{\pm}$; then we consider
\begin{align}
\ket{\Psi^{(2)}} &= r_1 \ket{2^+, +, +} +r_2 \ket{+, 2^+, +}+ r_3 \ket{+, +, 2^+} \nonumber \\
&+ \sqrt{2 r_1 r_2} \ket{1^+, 1^-, +} + \sqrt{2 r_1 r_3} \ket{1^+, -, 1^-} + \sqrt{2 r_2 r_3} \ket{+, 1^+, 1^-},
\end{align}
whose RDM is
\begin{align}
\rho_A^{(2)} =
\l
\begin{array}{c|cccccc}
&\ket{+,+} & \ket{+,1^-} & \ket{1^+,+} &\ket{2^+,+} & \ket{+,2^+} & \ket{1^+,1^-} \\
\hline
\bra{+,+} & r_2^2 & 0 & 0 & 0 & 0 & 0 \\
\bra{+,1^-} & 0 & 2 r_2 r_3 & 0 & 0 & 0 & 0\\
\bra{1^+,+} & 0 & 0 & 2 r_1 r_2 & 0 & 0 & 0\\
\bra{2^+,+} & 0 & 0 & 0 & r_1^2 & r_1 r_3 & 0 \\
\bra{+,2^+} & 0 & 0 & 0 & r_1 r_3 & r_3^2 & 0 \\
\bra{1^+,1^-} & 0 & 0 & 0 & 0 & 0 & 2 r_1 r_3 \\
\end{array}
\r.
\end{align}
In analogy with Section \ref{sec:two-kink}, a mixing occurs between the state with a pair kink-antikink in $A_1$ and that with the pair in $A_3$. For completeness, we diagonalise the RDM as
\be
\rho_A^{(2)} = r_2^2 \ket{\Phi_0^+}\bra{\Phi_0^+} + 2 r_1 r_2 \ket{\Phi_1^-}\bra{\Phi_1^-} + 2 r_2 r_3 \ket{\Phi_1^+}\bra{\Phi_1^+} + 2 r_1 r_3 \ket{\Phi_2^-}\bra{\Phi_2^-} + (r_1^2 + r_3^2) \ket{\Phi_2^+}\bra{\Phi_2^+},
\ee
with $\ket{\Phi_0^+} = \ket{+,+}$, $\ket{\Phi_1^-} = \ket{1^+,+}$, $\ket{\Phi_1^+}=\ket{+,1^-}$, $\ket{\Phi_2^-}=\ket{1^+,1^-}$ and
\be
\label{eq:Phi_2_eigenstate}
\ket{\Phi_2^+} = \frac{r_1 \ket{2^+,+} + r_3 \ket{+,2^+}}{\sqrt{r_1^2 + r_3^2}}.
\ee
We find, as expected, the same non-zero spectrum reported in Section \ref{sec:two-kink}.

Finally, we discuss the situation of a generic number $N$ of kinks. In analogy with the case of indistinguishable particles \cite{cdds-19a}, we consider an $N$-kink state
\begin{align}
\label{eq:N_kink_state_qubit}
\ket{\Psi^{(N)}} &= \underbrace{\sum_{\substack{N_1, N_2, N_3 \in \mathbb{N} \\ N_1 + N_2 + N_3 = N}} \sqrt{\frac{N! r_1^{N_1}r_2^{N_2}r_3^{N_3}}{N_1! N_2! N_3!}} \delta_{N_1 + N_2 + N_3, N} \ket{N_1^+, N_2^{(-)^{N_1}}, N_3^{(-)^{N_1+N_2}}}}_{\text{Kinks in all the three regions}} \nonumber \\
&+ \sum_{N_1=1}^{N-1}\sqrt{\binom{N}{N_1}r_1^{N_1}r_2^{N-N_1}} \ket{N_1^+, (N-N_1)^{(-)^{N_1}},(-)^N} \nonumber \\
&+ \sum_{N_1=1}^{N-1}\sqrt{\binom{N}{N_1}r_1^{N_1}r_3^{N-N_1}} \ket{N_1^+, (-)^{N_1},(N-N_1)^{(-)^{N_1}}} \nonumber \\
&+ \underbrace{\sum_{N_1=1}^{N-1}\sqrt{\binom{N}{N_1}r_2^{N_1}r_3^{N-N_1}} \ket{+, N_1^+,(N-N_1)^{(-)^{N_1}}}}_{\text{Kinks in two regions}} \nonumber \\
&+ \underbrace{r_1^{N/2}\ket{N, (-)^N,(-)^N}+r_2^{N/2}\ket{+, N^+, (-)^N}+r_3^{N/2}\ket{+, +, N^+}}_{\text{Kinks in one region}}.
\end{align}
In the expression above $\ket{N^{\pm}_j}$ is the state with $N_j$ kinks in a given region with the vacuum $\ket{\pm}$ on their left. The coefficients of the state are the square roots of the probabilities of finding a certain configuration of kinks among the regions $A_1$, $A_2$, $A_3$. While the RDM does not have a very insightful structure, we write its diagonal form as
\be
\label{eq:N_kinks_diagonal_RDM_qubit}
\rho_A^{(N)} = \lambda_0 \ket{\Phi_0}\bra{\Phi_0} + \sum_{\substack{\epsilon=\pm \\ N_A=1,\dots,N}} \lambda_{N_A}^\epsilon \ket{\Phi_{N_A}^\epsilon}\bra{\Phi_{N_A}^\epsilon}.
\ee
The eigenstates $\ket{\Phi_{N_A}^\epsilon}$ have a fixed number $N_A$ of kinks in $A_1 \cup A_3$ and a fixed parity $\epsilon$ of the number of kinks in one of these two regions: without loss of generality, we choose $A_1$, so $\epsilon =+$ iff there is an even number of kinks in $A_1$. The parity of the number of kinks/antikinks in $A_3$ is then automatically fixed by $N_A$ and $N$. In the eigenspace with $N_A$ and $\epsilon$ fixed, the qubit states are combined in a superposition to form $\ket{\Phi_{N_A}^\epsilon}$, as in Eq. \eqref{eq:Phi_2_eigenstate}. 
The mechanism behind the mixing is the following: two states form a linear superposition if and only if they have the same number $N_A$ \textit{and} they can be obtained one from the other by moving one magnon (i.e. a pair kink/antikink) from one region to the other. For every $N_A \ne 0$ there are two eigenspaces labeled by the parity $\epsilon$, while for $N_A = 0$ there is a single one-dimensional eigenspace. Specifically, the eigenvalues are:
\be
\label{eq:0_kinks_in_A_eigenvalue}
\lambda_{0} = r_2^N,
\ee
and for $N_A=1,\dots,N$
\begin{align}
\label{eq:N_A_kinks_in_A_eigenvalue}
\lambda_{N_A}^+ &= \sum_{j=0,
j \text{ even}}^{ N_A} \frac{N!}{j!(N_A - j)! (N-N_A)!}r_1^j r_3^{N_A - j}r_2^{N-N_A},\nonumber \\ \lambda_{N_A}^- &= \sum_{j=1, j \text{ odd}}^{ N_A} \frac{N!}{j!(N_A - j)! (N-N_A)!}r_1^j r_3^{N_A - j}r_2^{N-N_A},
\end{align}
where in every sum $j$, $N_A-j$ and $N-N_A$ correspond to the number kinks in $A_1$, $A_3$, $A_2$ respectively. 
From \eqref{eq:0_kinks_in_A_eigenvalue} and \eqref{eq:N_A_kinks_in_A_eigenvalue} we can finally express the tripartite $n$th R\'enyi entropy of the $N$-kink state as
\be
S_n = \frac{1}{1-n}\log\left[\lambda_{0}^n + \sum_{N_A = 1}^N \l(\lambda_{N_A}^+)^n + (\lambda_{N_A}^-)^n\r\right].
\ee
The above expressions can be checked via a direct calculation; in Appendix \ref{app:qubit_34_kinks}, as a detailed example, we discuss explicitly the cases with $N=3,4$.

\section{Field-theoretic approach}
\label{Sec:field-theoretic-approach}

In this section, we discuss the origin of the unusual entanglement content of kink excitations within a general field-theoretical framework. Our discussion relies on the description of those excitations in terms of semilocal fields acting on a symmetry-broken ground state, and their relations with the twist fields, which are the building blocks to compute entanglement measures via replica trick. In particular, we characterise the algebraic relations between the twist fields and semilocal fields such as the disorder operator in the Ising QFT. We recall that, as explained in Ref. \cite{cdmsc-22}, the algebraic relations between twist fields and local operators are sufficient to recover the entanglement content of quasi-particles; here, by generalising the commutation relations so to take into account the semilocality of the disorder operator, we are able to obtain the entanglement content of kink excitations.

We consider a 1+1 QFT displaying spontaneous symmetry breaking of a $\mathbb{Z}_2$ symmetry, for instance the Ising model in the ferromagnetic phase, and we denote its vacua by $\ket{\pm}$. We take a finite-size system of length $L$, and we specify the boundary conditions at the extremal points $x=0,L$. In particular, we denote by $b_{\pm}$ the boundary conditions associated with the presence of positive/negative boundary magnetic field. We assume that the asymptotic spectrum of the theory is described by kinks interpolating between the two vacua. Due to topological constraints, if the boundary conditions at $x=0,L$ are equal, say they are both $b_+$, the ground state is the vacuum $\ket{+}$ and kinks above it can only appear in an even number. On the other hand, if the system has different boundary conditions at the two edges, it can only host an odd number of kinks and the ground state of the finite-size system is a one-kink state.

Let $\mu(x)$ be a disorder field, that is a semilocal field connecting the two vacua in the region $x'>x$. Roughly speaking \footnote{
In the context of Integrable Field Theories \cite{yz-91,bb-92,dsc-96}, one usually refers to $\mu$ as a field with UV scaling dimension $\Delta_\mu = 1/8$, while the elementary excitations correspond to the fermionic field with dimension $\Delta = 1/2$; there is no direct relation between those fields, and $\mu$ has non-vanishing form factors with an arbitrary odd number of kinks. However, only the semilocal properties of the fields will enter our calculation and these issues do not play a role.}, $\mu(x)$ generates a kink at position $x$, and we are interested in the associated state at a given momentum. A single kink state can only connect sectors with different boundary conditions. For instance, the state
\be
\mu(x)\ket{+},
\ee
 interpolates between $b_+$ (at $x=0$) and $b_-$ (at $x=L$). We now construct the Fourier transform of $\mu(x)$ as
\be\label{eq:F_transf}
\mu(p) =  \int dx \ (e^{-ipx} +\mathcal{R}e^{ipx})\mu(x),
\ee
which generates a kink excitation at a given momentum (in absolute value). Here, $\mathcal{R}$ is a phase ($|\mathcal{R}|=1$) which depends on the boundary conditions of the field $\mu$, but it does not play a major role in the discussion below. We describe eigenstates with a finite number of kinks (in the large volume limit) as
\be\label{eq:exc_states}
\begin{cases}
\mu(p_1)\dots \mu(p_{2N})\ket{+}, \quad  (b_+,b_+) \text{ sector},\\
\mu(p_1)\dots \mu(p_{2N})\ket{-}, \quad  (b_-,b_-) \text{ sector},\\
\mu(p_1)\dots \mu(p_{2N+1})\ket{+}, \quad  (b_+,b_-) \text{ sector},\\
\mu(p_1)\dots \mu(p_{2N+1})\ket{-}, \quad  (b_-,b_+) \text{ sector}.
\end{cases}
\ee
We remark that, in principle, the momenta associated with eigenstates of the finite-size systems have to be quantised; however, this issue does not play a role as long as the number of particles is finite and the momenta are kept fixed in the large volume limit (as in Ref. \cite{cdmsc-22}).

Let us now define smeared fields by restricting the support of $\mu(p)$ to some spatial region $A$. That is, we introduce
\be
\mu_A(p) =  \int_A dx \ (e^{-ipx} +\mathcal{R}e^{ipx})\mu(x).
\ee
Then, we study the product of two smeared fields in the semiclassical limit where the momenta are fixed and the size of the regions becomes very large:
\be\label{eq:prod_mumud}
\begin{split}
\l\mu_{A'}(p') \r^\dagger \mu_A(p) = \int_A dx \int_{A'} dx' \ (e^{-ipx} +\mathcal{R}e^{ipx}) \ (e^{ip'x'} +\overline{\mathcal{R}}e^{-ip'x'}) \mu^\dagger(x') \mu(x) \simeq \\
\int_A dx \int_{A'} dx' \ (e^{-ipx} +\mathcal{R}e^{ipx}) \ (e^{ip'x'} +\overline{\mathcal{R}}e^{-ip'x'}) \la\mu^\dagger(x')\mu(x)\ra, 
\end{split}
\ee
where we only kept the lightest contribution coming from the fusion $\mu^\dagger \times \mu \rightarrow 1$, as all terms corresponding to heavier operators are less relevant in the limit of large regions; thus, we replace $\mu^\dagger(x') \mu(x)$ with its vacuum expectation value over $\ket{+}$ (a similar discussion can be found in Ref. \cite{cdmsc-22} with periodic boundary conditions for local fields). We can further simplify Eq. \eqref{eq:prod_mumud} if we assume that $\la\mu^\dagger(x)\mu(x')\ra$ decays fast enough with the distance $|x-x'|$, e.g. exponentially as it happens in the massive (ordered) phase of the Ising model. In that case, we can perform the change of variable $x' = x+x''$ and approximate
\be
\begin{split}
\l\mu_{A'}(p') \r^\dagger \mu_A(p) \simeq  \int dx'' \ \int_{A \cap A'}dx (e^{i(p'-p)x+ip'x''} + \mathcal{R}e^{i(p+p')x+ip'x''}+\text{c.c.})\la \mu^\dagger(x)\mu(x+x'')\ra \simeq\\
\int dx'' \l V_{A \cap A'}\delta_{p,p'}e^{ipx''} + V_{A \cap A'} \delta_{p+p',0}\mathcal{R} + \text{c.c.}\r \la \mu^\dagger(x)\mu(x+x'')\ra,
\end{split}
\ee
where the vacuum correlation function $\la \mu^\dagger(x)\mu(x+x'')\ra$ is assumed to be $x$-independent if $x$ is sufficiently distant from the boundary points and we denoted by $ V_{A \cap A'}$ the volume of the region $A \cap A'$. Eventually, we get
\be\label{eq:semiclass_contr}
\l\mu_{A'}(p') \r^\dagger \mu_A(p) \propto V_{A \cap A'}\delta_{p,p'},
\ee
where the proportionality constant comes from the integration of the vacuum correlation function. 

We remark that the considerations above hold only if the momentum is kept fixed in the large-volume limit. However, in some relevant situations, this is not the case. For example, the lowest-lying one-kink state is described by a single-body wave function $\sim \sin\l \pi x/L\r$ ($p=\pi/L$ and $\mathcal{R}=-1$ in Eq. \eqref{eq:F_transf}) if Dirichlet boundary conditions are present. In particular, the density of particles is not uniform in the large-volume limit, and the state looks inhomogeneous. These cases can be tackled as well with minor modifications. In particular, one can consider general smearing induced by a function $f(x)$ as
\be
\mu_f \equiv \int dx \ f(x)\mu(x).
\ee
Inhomogeneous states can then be studied by computing correlation functions of these smeared fields; however, this is beyond the purpose of our work.

\subsection{Twist fields and their algebra}\label{sec:tfield_algebra}

We first review some properties of the twist fields and their relations with the local operators, following Refs. \cite{cd-11,bd-17,cdmsc-22}, and then we generalise the discussion to the case of the semilocal disorder operator. We anticipate that the twist fields relevant to our discussion are those with an even number of disorder lines attached to them and associated with distinct copies of the replica field theory. To the best of our knowledge, these fields have been studied for the first time in our previous work \cite{cm-23} in the context of $\mathbb{Z}_2$ entanglement asymmetry; in particular, they have the same scaling dimension of the standard twist field but different monodromy properties, which become manifest in the ordered phase. We also mention that in previous works, for the symmetry-resolved entanglement, twist fields attached to a single disorder line have been studied in the disordered phase \cite{hc-20,cm-23a}: these are the \textit{composite branch-point twist fields} obtained from the fusion of the standard field with the disorder operator.

Let us consider $n$ replicas of the QFT. A field $\mathcal{O}(x)$ inserted in the $j$th replica is denoted by $\mathcal{O}^j(x)$. We focus on an algebra $\mathcal{A}$ of local fields; in the Ising model, this is the algebra generated by the order operator $\sigma(x)$ and the energy density $\varepsilon(x)$. To compute the entropy of a spatial region, that is the entropy of the subalgebra of $\mathcal{A}$ associated with that region, we employ the branch-point twist fields. These are fields in the replica model satisfying
\be\label{eq:def_twist}
\begin{cases}
\mathcal{T}(x)\mathcal{O}^j(y) = \mathcal{O}^{j}(y)\mathcal{T}(x), \quad x>y,\\
\mathcal{T}(x)\mathcal{O}^{j}(y) = \mathcal{O}^{j+1}(y)\mathcal{T}(x), \quad x<y.\\
\end{cases}
\ee
Loosely speaking, we say that $\mathcal{T}(x)$ introduces a branch cut at $x'>x$ which connects the replicas through the permutation $j\rightarrow j+1$. Similarly, we introduce a conjugate twist field $\tilde{\mathcal{T}}(x)$ such that
\be
\begin{cases}
\tilde{\mathcal{T}}(x)\mathcal{O}^j(y) = \mathcal{O}^{j}(y)\tilde{\mathcal{T}}(x), \quad x>y,\\
\tilde{\mathcal{T}}(x)\mathcal{O}^{j}(y) = \mathcal{O}^{j-1}(y)\tilde{\mathcal{T}}(x), \quad x<y\\
\end{cases}
\ee
acting as the permutation $j\rightarrow j-1$ on $x'>x$. $\mathcal{T}$ and $\tilde{\mathcal{T}}$ are the building blocks to construct the R\'enyi entropy. For example, one can express the moments of the RDM of $\ket{\Psi}$ associated with the region $A = [x_1,x_2]$ as
\be
\text{Tr}\l \rho^n_A\r \propto \ ^{n}\bra{\Psi}\mathcal{T}(x_1)\tilde{\mathcal{T}}(x_2)\ket{\Psi}^n.
\ee
Here $\ket{\Psi}^n \equiv \ket{\Psi}^{\otimes n}$ is the replicated state, and the non-universal proportionality constant can be absorbed in the normalisation of the twist fields.

\begin{figure}[t]
    \centering
	\includegraphics[width=0.9\linewidth]{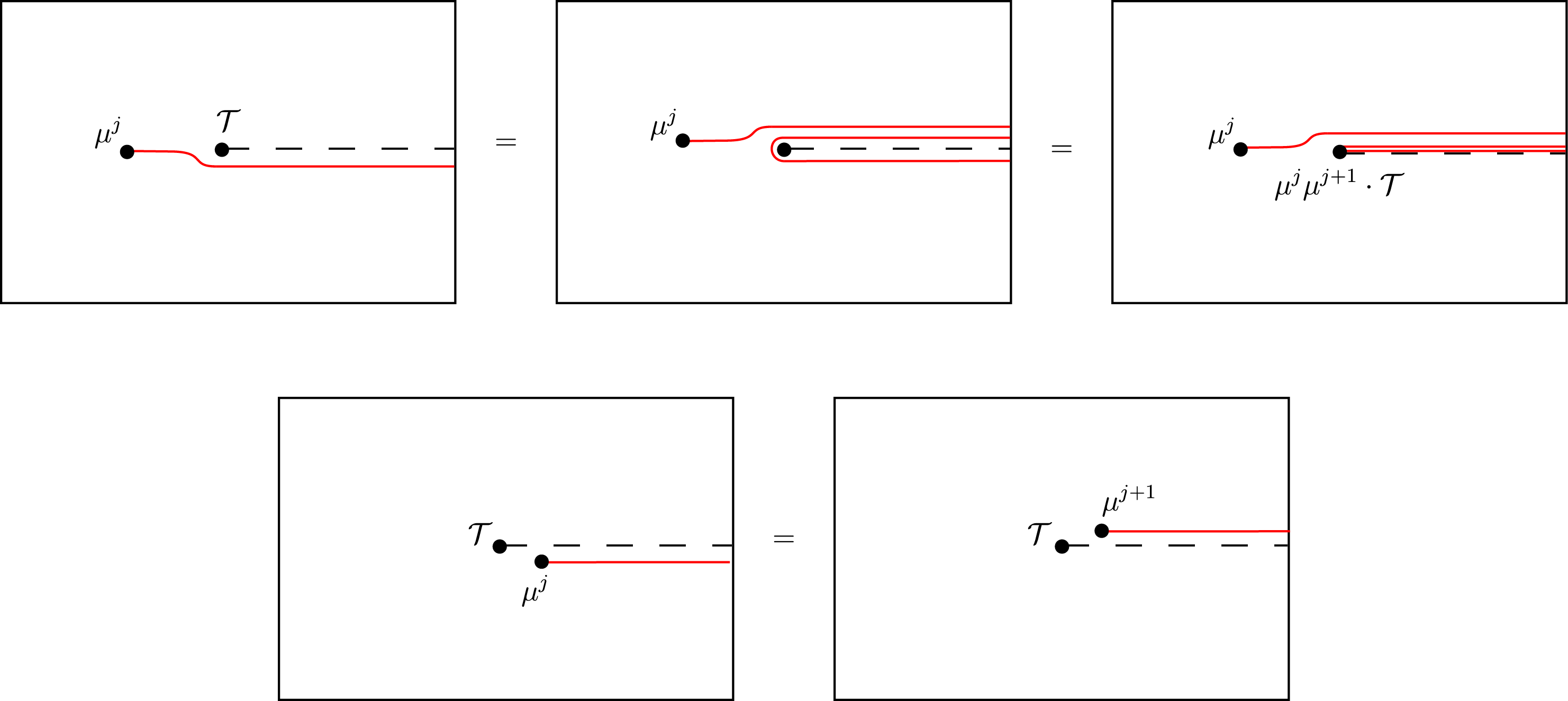}
    \caption{Commutation relations between the twist field $\mathcal{T}$ and the disordered operator in the $j$th replica ($\mu^j$). Top: The disorder line, in red, is deformed around the branch cut; the lower part of the line above crosses the branch cut, and it generates an additional line in the replica $j+1$. In this way, the operator $\mu^{j}\mu^{j+1}\cdot \mathcal{T}$ is obtained. Bottom: the whole disorder line is moved slightly above the branch cut and, yielding $\mu^{j+1}$.}
    \label{fig:com_rel}
\end{figure}

While Eq. \eqref{eq:def_twist} is sufficient to reconstruct the semiclassical predictions of the entropy of particles \cite{cdmsc-22} (generated by local operators), when kinks are involved one needs to understand the corresponding generalisation of Eq. \eqref{eq:def_twist} to semilocal operators. To the best of our knowledge, this problem has not been considered in the previous literature. We conjecture that the commutation relations between $\mu$ and $\mathcal{T}$ read as follows
\be\label{eq:Tfield_mu}
\begin{cases}
\mathcal{T}(x)\mathcal{\mu}^j(y) = \mathcal{\mu}^{j}(y)( \mu^{j}\mu^{j+1} \cdot \mathcal{T} )(x), \quad x>y,\\
\mathcal{T}(x)\mathcal{\mu}^{j}(y) = \mathcal{\mu}^{j+1}(y)\mathcal{T}(x), \quad x<y,\\
\end{cases}
\ee
with $(\mu^{j}\mu^{j+1} \cdot \mathcal{T})(x)$ the lightest field generated by the fusion
\be
   \mu^{j}\mu^{j+1}\times \mathcal{T},
\ee
and $(\mu^{j}\mu^{j+1})(x)$ is a semilocal field of the replica model with disorder lines on replicas $j$ and $j+1$. Physically, this means that $\mu$ charges the standard twist field $\mathcal{T}$ generating a new composite field $\mu^{j}\mu^{j+1} \cdot \mathcal{T}$. We give a pictorial representation of Eq. \eqref{eq:Tfield_mu} in Fig. \ref{fig:com_rel}: the branch cut of the twist field is represented as a black dashed line while the disorder line is the red line.

A simple derivation of Eq. \eqref{eq:Tfield_mu} can be obtained assuming that $\mu(x)$ is generated by a product of local spin flip operators $\mathcal{O}(x')$ inserted at position $x'>x$
\be
\mu(x) = \prod_{x'>x}\mathcal{O}(x).
\ee
In the case $x>y$, a straightforward computation gives
\be
\begin{split}
\mathcal{T}(x)\mu^j(y) = &\mathcal{T}(x)\prod_{y'>y}\mathcal{O}^j(y') = \prod_{y<y'<x}\mathcal{O}^j(y') \prod_{y'>x}\mathcal{O}^{j+1}(y')\mathcal{T}(x)=\\
&\prod_{y'>y}\mathcal{O}^j(y')\prod_{y'>x}\mathcal{O}^j(y')\prod_{y'>x}\mathcal{O}^{j+1}(y')\mathcal{T}(x) = \mu^j(y)\mu^j(x)\mu^{j+1}(x)\mathcal{T}(x),
\end{split}
\ee
where the locality of $\mathcal{O}(x)$ has been employed, through Eq. \eqref{eq:def_twist}, together with the property $\mathcal{O}(x)\mathcal{O}(x)=1$. We remark that, while the derivation above can be made rigorous in the lattice, its extension to field theories is not straightforward due to divergences coming from the insertion of fields at the same points; we claim that these issues are only technical and Eq. \eqref{eq:Tfield_mu} still holds in the QFT.

In general, the commutation relation between $\mathcal{T}$ and the string $(\mu^{j_1}\dots \mu^{j_k})(x)$, associated with disorder lines in the replicas $j_1\dots,j_k$, can be found in a similar way: each disorder line $\mu^j$ crosses the branch cut, and it charges the twist field with an additional insertion of $\mu^{j}$ and $\mu^{j+1}$. Furthermore, two disorder lines associated with the same replica simplify (due to the $\mathbb{Z}_2$ fusion rule $\mu \times \mu \rightarrow 1$). In conclusion, via this procedure only twist fields with an even number of disorder lines attached to them are generated, and we refer the interested reader to Ref. \cite{cm-23} for the characterisation of their form factors. For our purposes, we only need to recall the properties of the 0-kink form factors, namely the matrix elements between the ($2^n$) vacua in the replica theory. In particular, due to the topological constraints, it was argued in \cite{cm-23} that the only non-vanishing 0-kink form factors for the operators above are
\be
^{n}\bra{+}\mathcal{T}(x)\ket{+}^{n} = \ {}^{n}\bra{-}\mathcal{T}(x)\ket{-}^{n} \propto m^{-\Delta_\mathcal{T}}.
\ee
Here, $m^{-1}$ is a correlation length, $\Delta_\mathcal{T}$ is the scaling dimension of $\mathcal{T}$, and the non-universal proportionality constant depends on the normalisation of the field. A technical observation is in order: these results refer to the theory at infinite size, while here we are interested in the finite-size counterpart. However, as investigated in Ref. \cite{cdds-18b}, we expect deviations from the infinite-size theory to be exponentially suppressed in the system size, and therefore we neglect those in the forthcoming discussion.

\subsection{One-kink entropy}
The algebraic relations \eqref{eq:Tfield_mu} summarise the essential distinction between (local) particles and kink excitations at the level of entropy of regions. We now claim that these relations are sufficient to give explicit predictions for the R\'enyi entropies in the semiclassical limit where the regions are large compared to the microscopic lengths. In this section we provide an application of the formalism developed so far by analysing a simple yet non-trivial case, that is the entropy of a single kink in a tripartite geometry. 

We consider an interval $A = [\ell_1,\ell_2]$ and its complement $B = B_1\cup B_2 = [0,\ell_1]\cup[\ell_2,L]$ and we study the R\'enyi entropy of the state
\be
\ket{\Psi} = \mu(p)\ket{+}.
\ee
To do so, we need to compute the ratio
\be\label{eq:exp_valueT}
\frac{^{n}\bra{\Psi}\mathcal{T}(\ell_1)\tilde{\mathcal{T}}(\ell_2)\ket{\Psi}^n}{^{n}\braket{\Psi|\Psi}^n},
\ee
where the denominator ensures proper normalisation, and compare it with
\be
^{n}\bra{+}\mathcal{T}(\ell_1)\tilde{\mathcal{T}}(\ell_2)\ket{+}^n,
\ee
which yields the ground state contribution. Let us first compute the normalisation $^{n}\braket{\Psi|\Psi}^n$ as follows
\be
^{n}\braket{\Psi|\Psi}^n = \l\braket{\Psi|\Psi}\r^n = \l\bra{+}(\mu(p))^\dagger \mu(p)\ket{+}\r^n \propto L^n,
\ee
where Eq. \eqref{eq:semiclass_contr} and $\braket{+|+}=1$ have been employed.

\begin{figure}[t!]
    \centering
	\includegraphics[width=\linewidth]{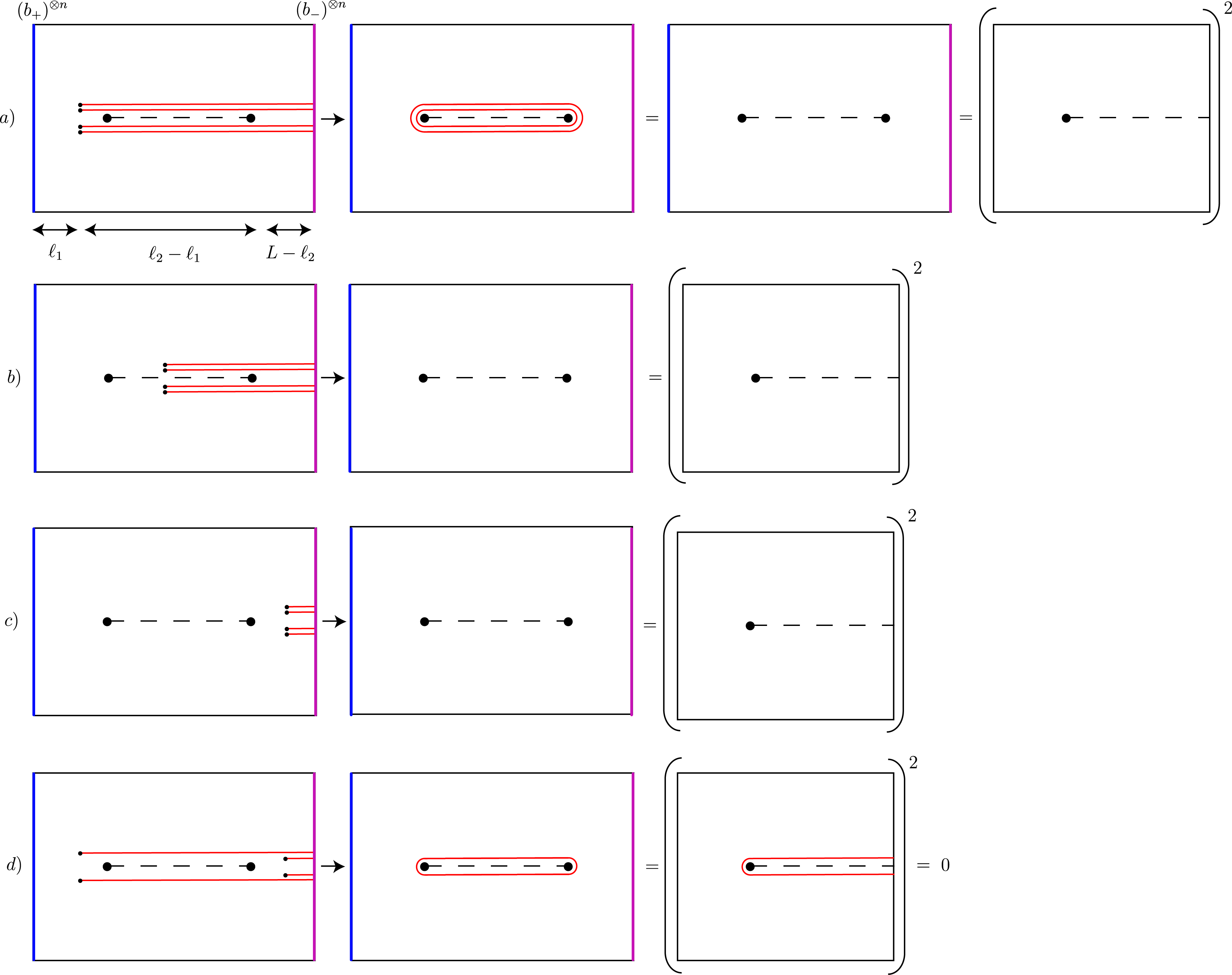}
    \caption{
    Schematic structure of the correlation functions between disorder operators and twist fields in the $n$th replica model ($n=2$). The blue/pink lines correspond to the left/right boundary points ($x=0,L$) associated with boundary conditions $(b_+)^{\otimes n}$ and $(b_-)^{\otimes n}$ respectively. The dashed black line is the branch cut and the disorder lines are depicted in red.
    a) Contribution coming from $\mu^1$ and $\mu^2$ inserted in $B_1 = [0,\ell_1]$. Here, after the contraction the disorder lines cancel each other and they give rise to $^2\bra{+}\mathcal{T}(\ell_2)\tilde{\mathcal{T}}(\ell_2) \ket{+}^2$. b) Contribution from $\mu^1$ and $\mu^2$ inserted at $A = [\ell_1,\ell_2]$. c) Contribution from $\mu^1$ and $\mu^2$ inserted at $B_2 = [\ell_2,L]$. d) Vanishing term associated with the insertion of $\mu^1$ in $B_1$ and $\mu^2$ in $B_2$. Here, a pair of composite twist fields is generated after the contraction and the corresponding expectation value is zero (in the large volume limit).}
    \label{fig:Renyi_contractions}
\end{figure}

To proceed with the computation, it is convenient to split the field $\mu^j(p)$ in terms of its spatial restrictions:
\be
\mu^j(p) = \mu^j_{A}(p)+\mu^j_{B_1}(p)+\mu^j_{B_2}(p).
\ee
The operators above have definite commutation relations with the twist fields, which, as a consequence of \eqref{eq:Tfield_mu}, read
\be\label{eq:Tfield_mu1}
\begin{split}
\mathcal{T}(\ell_1)\tilde{\mathcal{T}}(\ell_2)\mu^j_{B_1}(p) &= \mu^j_{B_1}(p)(\mu^j\mu^{j+1}\cdot \mathcal{T} )(\ell_1)( \mu^j\mu^{j+1} \cdot \tilde{\mathcal{T}} )(\ell_2),\\
\mathcal{T}(\ell_1)\tilde{\mathcal{T}}(\ell_2)\mu^j_{B_2}(p) &= \mu^j_{B_2}(p)\mathcal{T}(\ell_1)\tilde{\mathcal{T}}(\ell_2),\\
\mathcal{T}(\ell_1)\tilde{\mathcal{T}}(\ell_2)\mu^j_{A}(p) &= \mu^{j+1}_{A}(p)\mathcal{T}(\ell_1)( \mu^j\mu^{j+1} \cdot \tilde{\mathcal{T}})(\ell_2).
\end{split}
\ee
We expand the numerator of Eq. \eqref{eq:exp_valueT} in terms of the restrictions of the fields as
\be\label{eq:eq:exp_valueT1}
\begin{split}
^{n}\bra{\Psi}\mathcal{T}(\ell_1)\tilde{\mathcal{T}}(\ell_2)\ket{\Psi}^n &= {}^n\bra{+} (\mu^n(p))^\dagger\dots(\mu^1(p))^\dagger
 \mathcal{T}(\ell_1)\tilde{\mathcal{T}}(\ell_2) \mu^1(p)\dots\mu^n(p)\ket{+}^n \\&=
{}^n\bra{+} (\mu^n(p))^\dagger\dots(\mu^1(p))^\dagger
 \mathcal{T}(\ell_1)\tilde{\mathcal{T}}(\ell_2) \\&\times (\mu^1_A(p)+\mu^1_{B_1}(p)+\mu^1_{B_2}(p))\dots(\mu^n_A(p)+\mu^n_{B_1}(p)+\mu^n_{B_2}(p))\ket{+}^n.
\end{split}
\ee
We employ Eq. \eqref{eq:Tfield_mu1}, bringing the restrictions of $\mu^j(p)$ to the left of the twist fields, so that they can be eventually contracted with $\l\mu^j(p)\r^\dagger$. The contractions produce a total of $3^n$ terms. However, most of these are vanishing in the limit we are considering 
For instance, since Eq. \eqref{eq:semiclass_contr} gives
\be
(\mu^{j'}(p))^\dagger \mu^j_{C}(p) \propto V_C \delta_{jj'},
\ee
for any region $C$, it follows that only contractions between disorder lines in the same replica survive. In addition, among the remaining terms, we should only keep those for which the field obtained after performing all the commutation relations is the standard twist field, otherwise, the corresponding vacuum expectation values vanish. 
In conclusion, only three terms remain from the expansion in Eq. \eqref{eq:eq:exp_valueT1}, so we can write
\be
\begin{split}
^{n}\bra{\Psi}\mathcal{T}(\ell_1)\tilde{\mathcal{T}}(\ell_2)\ket{\Psi}^n \simeq \  &{}^n\bra{+} (\mu^n(p))^\dagger\dots(\mu^1(p))^\dagger
\mu^1_{B_1}(p)\dots\mu^n_{B_1}(p) \mathcal{T}(\ell_1)\tilde{\mathcal{T}}(\ell_2) \ket{+}^n+\\
&{}^n\bra{+} (\mu^n(p))^\dagger\dots(\mu^1(p))^\dagger
\mu^1_{B_2}(p)\dots\mu^n_{B_2}(p) \mathcal{T}(\ell_1)\tilde{\mathcal{T}}(\ell_2) \ket{+}^n+\\
&{}^n\bra{+} (\mu^n(p))^\dagger\dots(\mu^1(p))^\dagger
\mu^2_{A}(p)\dots\mu^1_{A}(p) \mathcal{T}(\ell_1)\tilde{\mathcal{T}}(\ell_2) \ket{+}^n \propto\\
& (\ell_1^n + (\ell_2-\ell_1)^n + (L-\ell_2)^n)\times{}^{n}\bra{+}\mathcal{T}(\ell_1)\tilde{\mathcal{T}}(\ell_2)\ket{+}^n.
\end{split}
\ee
A schematic representation of the mechanism depicted above is sketched in Fig. \ref{fig:Renyi_contractions}.

Putting everything together, we compute the following universal ratio
\be
\frac{^{n}\bra{\Psi}\mathcal{T}(\ell_1)\tilde{\mathcal{T}}(\ell_2)\ket{\Psi}^n}{{}^n\braket{\Psi|\Psi}^n\times ^{n}\bra{+}\mathcal{T}(\ell_1)\tilde{\mathcal{T}}(\ell_2)\ket{+}^n} \simeq (r_1^n+r^n_2+r_3^n),
\ee
with $r_j$ defined by
\be
r_1 \equiv \frac{\ell_1}{L}, \quad r_2 \equiv \frac{\ell_2-\ell_1}{L}, \quad r_3 \equiv \frac{L-\ell_2}{L},
\ee
corresponding to the probabilities to find the kink in the region $B_1,A,B_2$ respectively. We finally express the difference of entropy between $\ket{\Psi}$ and $\ket{+}$ as
\be\label{eq:sn_1_kink}
S_n-S_{n,0} = \frac{1}{1-n}\log\l r^n_1+r^n_2+r^n_3\r.
\ee
This is the same result \eqref{eq:entropy_1K_tripartite_lattice}, up to the vacuum entropy $S_{n,0}$ that is, in general, non-vanishing in a field theory. Remarkably, we find universality of the entropy difference between the low-lying excited states and the vacuum, which is one of the significant achievements of the field-theoretic framework.

Some comments regarding the regime of small regions are needed. So far, we analysed the scaling limit where the ratios of the sizes are fixed, and they are much larger than the microscopic lengths: in particular, the regime $\ell_2-\ell_1 \lesssim m^{-1}$ is not described by the previous formula. Indeed, formally, if $\ell_2=\ell_1$ the region $A$ becomes the empty set and the entropy vanishes. Unfortunately, this behavior is not recovered by the limit $r_2\rightarrow 0$ of Eq. \eqref{eq:sn_1_kink}, and it is not obvious a priori why there should be a residual entropy when the region $A$ is much smaller than the rest of the system. The reason is that, albeit the probability that the kink belongs to $A$ vanishes in the limit above, one can still distinguish whether the latter belongs to $B_1$ or $B_2$ just by measuring the local magnetisation in $A$. In particular, the state becomes locally indistinguishable from the statistical mixture 
\be
\la \dots\ra = r_1 \bra{+}\dots\ket{+} + r_3 \bra{-}\dots\ket{-},
\ee
and thus one observes an entropy difference $S_n-S_{n,0} = \frac{1}{1-n}\log\l r^n_1+r^n_3\r$.

\section{Kramers-Wannier duality and the entanglement of algebras} 
\label{Sec:Kramers-Wannier and entanglement of algebras}

In the previous sections, we have shown that the entanglement content of kinks explicitly differs from that of particles. This might sound surprising at first, since the ordered/disordered phases are related by duality in the quantum Ising chain: in particular, the kinks in the ordered phase are dual to the particles in the disordered phase. Naively, one could expect that the entanglement content of the corresponding excitations should be the same, but this is not the case. The only reasonable conclusion is that the entanglement of regions is not self-dual under Kramers-Wannier duality. This fact can be understood in terms of the non-local nature of the duality. In particular, it has been shown in \cite{dh-23} that the entanglement is transferred from local to non-local degrees of freedom. However, while it is well-established that some local and semilocal operators are related by duality (e.g. the order and disorder operators), the identification of a possible dual of twist fields or of the entropy of a region is less obvious. In this Section, we adopt an algebraic approach, in the spirit of Refs. \cite{op-04,chr-13,ch-22,Witten-18,lr-20}, to answer those questions. The formalism, based on the theory of (finite-dimensional) $C^{*}$-algebras, allows dealing systematically with observables that are not local in the spin basis, such as the disorder operators and provides a natural framework to properly define their entanglement content.

We give first a brief review of the Kramers-Wannier duality in the context of the quantum Ising chain. Then, we discuss the notion of density matrix associated with observables in the context of $C^{*}$-algebras. Finally, we identify the notion of the duals of twist fields and R\'enyi entropies associated with regions.

\subsection{Kramers-Wannier duality in a quantum spin-$1/2$ chain}\label{sec:KW_duality}

The Kramers-Wannier duality, in the context of quantum chains, is associated with the possibility of representing the states of a system using spin or kink variables. Specifically, this proves beneficial for models such as the Ising spin chain, as the Hamiltonian remains local in both representations, albeit ordered phases correspond to disordered phases.
Here, we are interested in a systematic treatment of the \textit{duality map} between the two descriptions. While this is usually non-invertible, and the Kramers-Wannier duality is known as a non-invertible symmetry \cite{amf-16}, specific sectors associated with given boundary conditions are in one-to-one correspondence as we explain below.

Let us consider the Hilbert space $\mathcal{H}$ generated by the $2^L$ configurations of spins  $\pm$ along the $z$ axis, for a chain of length $L$. This was defined in Eq. \eqref{eq:spin_one_half_Hilbert_space}. We call $\mathcal{H}$ the \textit{site space}, and the configurations of site variables give rise to an orthonormal basis. We consider another Hilbert space $\mathcal{H}'$ of dimension $2^{L-1}$, defined similarly:
\be
\mathcal{H}'= \text{Span}\{ \ket{s'_{3/2},\dots, s'_{L-1/2}}, \quad s'_{j+1/2} = \pm \}.
\ee
This is referred to as the \textit{bond space},  being generated by configurations of $L-1$ bond variables. We can associate any site configuration with a bond configuration considering the sign changes of adjacent sites. More precisely, we construct a linear map
\be
T: \mathcal{H} \rightarrow \mathcal{H}',
\ee
acting on the basis elements as follows
\be
T:\ket{s_1,\dots, s_L} \rightarrow  \ket{s'_{3/2},\dots, s'_{L-1/2}}, \quad s'_{j+1/2} = s_js_{j+1}.
\ee
Roughly speaking, we say that a kink at position $j+1/2$ is present if there is a change of sign between the sites $j$ and $j+1$. In the following, we refer to $T$ as the duality map.

We can also consider the adjoint of $T$ as
\be
T^\dagger: \mathcal{H}' \rightarrow \mathcal{H},
\ee
such that the associated matrices of $T$, $T^\dagger$ are hermitian conjugate of each other in the configuration basis. From the definition, it holds $T T^\dagger =1$, namely $T^\dagger$ is an isometry of $\mathcal{H}'$ onto itself. However, $T^\dagger T\neq 1$; this is not surprising, since the two Hilbert spaces $\mathcal{H},\mathcal{H}'$ have different dimensions, and, in particular, $T$ has to display a non-trivial kernel. We fix this issue by focusing on a sector of $\mathcal{H}$ such that the associated restriction of $T$ becomes invertible. We choose to fix the last spin and define two sectors of dimension $2^{L-1}$
\be\label{eq:Hpm}
\mathcal{H}_\pm \equiv \text{Span}\{ \ket{ s_1,\dots, s_{L-1},\pm}, \quad s_j = \pm \},
\ee
satisfying
\be
\mathcal{H} = \mathcal{H}_+ \oplus \mathcal{H}_-.
\ee
From now on we focus on $\mathcal{H}_+$ and, with a slight abuse of notation, we refer to $T$ as the restriction
\be
T:\mathcal{H}_+ \rightarrow \mathcal{H}'.
\ee
In this way, $T$ becomes invertible and both $T^\dagger T$ and $TT^\dagger$ give the identity operator (on $\mathcal{H}_+$ and $\mathcal{H}'$ respectively). 

The duality map we defined above is a correspondence between states,but  we can extend it to a map between observables. For instance, we consider
\be\label{eq:isom}
\begin{split}
\text{End}(\mathcal{H}_+) &\rightarrow \text{End}(\mathcal{H}')\\
\mathcal{O} &\mapsto \mathcal{O}' \equiv T\mathcal{O}T^\dagger.
\end{split}
\ee
This is an isomorphism between $C^{*}$-algebras, since it is invertible and $(\mathcal{O}_1\mathcal{O}_2)' = \mathcal{O}'_1\mathcal{O}_2'$, $(\mathcal{O}^\dagger)' = (\mathcal{O}')^\dagger$ hold. One can show, by checking, for example, the action of the operators on basis elements related by duality, that the following relations holds
\be\label{eq:isom1}
\begin{split}
\sigma^z_{j}\sigma^z_{j+1} &\mapsto \sigma^z_{j+1/2},\\
\sigma^x_{j} &\mapsto \sigma^x_{j-1/2}
\sigma^x_{j+1/2}.
\end{split}
\ee
Moreover, by representing the one-site operators as products of two-site operators and using the fact that Eq. \eqref{eq:isom} is an isomorphism, one gets
\be
\begin{split}
\sigma^z_{j} &\mapsto \prod_{j'\geq j}\sigma^z_{j'+1/2},\\
\prod_{j'\geq j}\sigma^x_{j'} &\mapsto \sigma^x_{j+1/2}.
\end{split}
\ee
This means in particular that local and semilocal operators are mixed with each other by the duality map. It is also worth stressing that, since $\sigma^z_j,\sigma^x_j$ generate the algebra $\text{End}(\mathcal{H}_{+})$ (via products and linear combinations), the relations above unambiguously characterise the isomorphism \eqref{eq:isom}.

Finally, we comment on some consequences of the non-locality of the duality map. Let us consider a spatial bipartition $\mathcal{H}_{+} =\mathcal{H}_{A}\otimes \mathcal{H}_{B}$ with
\be
\begin{split}
\mathcal{H}_{A} &= \text{Span}\{ \ket{s_1,\dots, s_\ell}, \quad s_j = \pm \},\\
\mathcal{H}_{B} &= \text{Span}\{ \ket{s_{\ell+1},\dots, s_{L-1}+}, \quad s_j = \pm \}.
\end{split}
\ee
Here, $A$ and $B$ are two subregions consisting of $\ell$ and $L-\ell$ sites respectively. Unfortunately there is no way to define the image of $\mathcal{H}_A$ under duality due to the non-locality of the latter: the map $T$ cannot be canonically extended to a map $T:\mathcal{H}_A\rightarrow \mathcal{H}'_A$ obtained from a possible spatial bipartition $\mathcal{H}' = \mathcal{H}'_A \otimes \mathcal{H}'_B$. Nonetheless, it is meaningful to map the observables associated with $A$ to other observables of $\mathcal{H}'$. Namely, the former can be first canonically embedded in the space of the observables of $\mathcal{H}_+$ via
\be
\text{End}(\mathcal{H}_A) \rightarrow \text{End}(\mathcal{H}_A)\otimes \mathds{1}_{\mathcal{H}_B} \subset \text{End}(\mathcal{H}_+),
\ee
and then the duality map \eqref{eq:isom} can be applied as
\be
\text{End}(\mathcal{H}_A)\otimes \mathds{1}_{\mathcal{H}_B} \rightarrow \text{End}(\mathcal{H}').
\ee
The price to pay is that the image of the map above contains observables, as it happens for $\sigma^z_j$ (with $j\leq \ell$) in Eq.\eqref{eq:isom1}, with a \text{string} (product) of $\sigma^z$ attached to them. In this respect, we say that the isomorphism \eqref{eq:isom} mixes local and semilocal observables.

\subsection{An algebraic definition of the reduced density matrix}\label{sec:dens_mat}

Here, we review how to construct the reduced density matrix associated with an algebra. This construction, which is well-established in the context of mathematical physics (see e.g. the textbook \cite{op-04}), is nonetheless mostly overlooked by the vast majority of works on entanglement. For instance, people usually refer to a spatial bipartition of a Hilbert space \cite{afov-08}
\be\label{eq:bipartition}
\mathcal{H} = \mathcal{H}_A \otimes \mathcal{H}_B,
\ee
and the entanglement of $A$ is probed via the properties of the reduced density matrix $\rho_A \in \text{End}(\mathcal{H}_A)$. This approach is satisfactory when the observables of interests are realised locally in the corresponding model. However, in the presence of semilocal operators, or when dealing with maps (such as the Kramers-Wannier duality) that do not preserve locality, a more general approach is needed. 

In the following, we will assume that the algebra of observables $\mathcal{A}$ is a finite-dimensional $C^{*}$-algebra. A state is then a linear functional from $\mathcal{A}$ to the complex numbers:
\be
\begin{split}
\mathcal{A} &\rightarrow \mathbb{C}\\
a &\mapsto \la a\ra,
\end{split}
\ee
and it corresponds to the usual notion of \textit{expectation value} of the observables in $\mathcal{A}$. As we show below, it is possible to construct a density matrix $\rho_{\mathcal{A}}$ as an observable itself requiring
\be
\text{Tr}\l \rho_{\mathcal{A}} a\r = \la a\ra, \quad \forall a \in \mathcal{A}.
\ee
The trace $\text{Tr}\l \dots\r$ entering the above expression is defined unambiguously thanks to the isomorphism between $\mathcal{A}$ and a direct sum of matrix algebras (briefly reviewed in Appendix \ref{app:cstar}), that is a standard result of the representation theory of $C^{*}$-algebras (see e.g. Ref. \cite{Landsman-98}). Specifically, given an orthonormal basis $\{a_i\}$ of $\mathcal{A}$
\be
\text{Tr}\l a^\dagger_i a_j\r = \delta_{ij},
\ee
one expresses $\rho_{\mathcal{A}}$ as
\be
\rho_{\mathcal{A}} = \sum_j r_j a^\dagger_j, \quad r_j \equiv \la a_j\ra.
\ee
By applying the definitions above to the bipartition \eqref{eq:bipartition} and the algebra $\text{End}\l \mathcal{H}_A\r\otimes \mathds{1}_{\mathcal{H}_B}  \subset \text{End}(\mathcal{H})$, one obtains
\be
\rho_{\mathcal{A}} = \rho_A \otimes \frac{\mathds{1}_{\mathcal{H}_B}}{\text{Tr}\l \mathds{1}_{\mathcal{H}_B}\r}.
\ee
The main advantage of this formulation is that $\rho_{\mathcal{A}}$ becomes part of the local algebra $\mathcal{A}$ and the local Hilbert space $\mathcal{H}_A$ is never explicitly used.

In the framework depicted above, it is possible to make sense of the dual of a density matrix under the Kramers-Wannier map considered in the previous Sec. \ref{sec:KW_duality}. Indeed, one can pick the algebra $\mathcal{A} \subseteq \text{End}(\mathcal{H}_+)$ associated with a region of the site space defined in Eq. \eqref{eq:Hpm}. Then, given a state $\la \dots\ra$ for $\mathcal{A}$ and its density matrix $\rho_{\mathcal{A}}$, from the map \eqref{eq:isom} one can define a \textit{dual density matrix}
\be\label{eq:dual_dmat}
\rho_{\mathcal{A}'} \equiv T \rho_{\mathcal{A}} T^\dagger \subset \text{End}(\mathcal{H}').
\ee
It is not difficult to show that $\rho_{\mathcal{A}'}$ is just the density matrix of the dual algebra
\be\label{eq:Aprime}
\mathcal{A}' \equiv T\mathcal{A}T^\dagger,
\ee
associated with the dual state defined by
\be\label{eq:state_prime}
\la TaT^\dagger\ra' \equiv \la a\ra, \quad a \in \mathcal{A}.
\ee
Indeed, by making use of the property $T^\dagger T=1$, a straightforward calculation yields
\be
\text{Tr}\l \rho_{\mathcal{A}'} a'\r = \text{Tr}\l \rho_{\mathcal{A}} a\r = \la a \ra = \la a'\ra', \quad a \in \mathcal{A},
\ee
with $a' = TaT^\dagger$ a generic observable in $\mathcal{A}'$, that is the defining property of the density matrix. It is worth pointing out that expressions such as $T\rho_A T^\dagger$, with $\rho_A \in \text{End}\l \mathcal{H}_A\r$, are meaningless since $T$ is defined at the level of the global Hilbert space only, which is the main reason we adopted the algebraic approach to investigate the duality.

Once the density matrices are defined in the algebraic framework, the usual entanglement measures can be introduced as well. For instance, the R\'enyi entropy is
\be
S_n = \frac{1}{1-n}\log \text{Tr}\l\rho^n_\mathcal{A}\r,
\ee
and it matches the standard definition (in Ref. \cite{afov-08}) if applied to the algebra of regions. Moreover, since $T$ is an invertible isometry, from Eq. \eqref{eq:dual_dmat} we get
\be\label{eq:equal_entropy}
\text{Tr}\l \rho_{\mathcal{A}}^n \r = \text{Tr}\l \rho_{\mathcal{A}'}^n \r,
\ee
and the two algebras have the same entropy for the corresponding dual states. At this point, it is worth giving a simple paradigmatic example to discuss the consequences of the construction above. We consider a state $\la \dots\ra$ of $\mathcal{H}$, defined in Eq. \eqref{eq:spin_one_half_Hilbert_space}, namely the ground state of an Ising chain in its ordered phase. We focus on a region $A$, we compute the associated entropy, that is the entropy of the algebra $\text{End}(\mathcal{H}_A)$, then we turn our attention to the Kramers-Wannier duality. For instance, we consider the dual state $\la \dots\ra'$, which is a ground state of the paramagnetic phase, and we aim to characterise the entropy of some region in this state. The main problem is that, albeit the states $\la \dots\ra$ and $\la \dots \ra'$ are related by duality, the algebras of regions are not. In particular, the local observables in the region $A$ are now mapped onto an algebra that is not associated with a region of the system $\mathcal{H}'$. Similar considerations hold for low-lying excited states with a finite number of kinks, dual to states with a finite number of particles.

In conclusion, the Kramers-Wannier duality does not give any direct relation regarding the entropy of regions. Nonetheless, we mention for completeness that, since the duality can be implemented by a Matrix Product Operator (MPO) of bond dimension $2$ \cite{ldov-23}, the entropy difference between a state and its dual has an upper bound given by $b \log 2$, with $b$ the number of entangling points.

\subsection{The dual of the twist field}\label{sec:d_tfield}

In this section, we discuss the twist operators in the algebraic framework and their transformation under Kramers-Wannier duality. In the context of spin chains, these operators have been defined in Ref. \cite{cd-11} via their explicit basis representation (see also Ref. \cite{cm-23}). Here, as we did for the reduced density matrix, we follow a slightly more abstract construction based on algebraic properties of finite dimensional $C^{*}$-algebras, relaxing the hypothesis of the strict locality of the observables in the computational basis. In this way, we can safely discuss the notion of twist operators associated with algebras and their dual.

Let us consider an algebra $\mathcal{A}$, we replicate it $n$ times and we call $\mathcal{A}^{\otimes n}$ the \textit{replica algebra}. $\mathcal{A}^{\otimes n}$ is generated by the operators $a^j$, obtained by inserting the elements $a \in \mathcal{A}$ in the $j$th replica, for every $j$:
\be
a^j \equiv 1\otimes \dots \otimes a\otimes 1 \otimes \dots 1.
\ee
The twist operator $\mathcal{T}_\mathcal{A} \in \mathcal{A}^{\otimes n}$ is an operator which acts as a replica cyclic permutation $j\rightarrow j+1$. More precisely, one requires
\be\label{eq:Twist_op_def}
\mathcal{T}_\mathcal{A} a^{j} = a^{j+1} \mathcal{T}_\mathcal{A}.
\ee 
Eq. \eqref{eq:Twist_op_def} is already present in \cite{cd-11}, where it is obtained from a more fundamental definition. Instead, here we ask whether this can be truly considered a defining property of the twist field. Clearly, if $\mathcal{T}_\mathcal{A}$ satisfies Eq. \eqref{eq:Twist_op_def}, then also $\lambda \mathcal{T}_\mathcal{A}$ with $\lambda \in \mathbb{C}$ has the same property, and one may wonder whether \eqref{eq:Twist_op_def} fixes the twist operator up to a proportionality constant.

In the discussion below, we assume that the algebra $\mathcal{A}$ is isomorphic to $\text{End}(\mathbb{C}^d)$ for a given $d$; this is the framework of main interest since it corresponds to both the algebra of regions in spin chains, and the (non-local) algebra arising from its Kramers-Wannier dual. In those cases, the only element of $\mathcal{A}$ commuting with every other element is, up to a constant, the identity $1\in \mathcal{A}$, and the center of the algebra is trivial. A few minor issues arise for other (finite-dimensional) algebras, and they are discussed in the Appendix \ref{app:cstar}.

The existence of a non-zero operator $\mathcal{T}_\mathcal{A}$ is not obvious a priori (we comment on a general algebraic proof in Appendix \ref{app:cstar}). However, since $\mathcal{A}$ can be realised as an algebra of matrices, one can employ the explicit construction of Ref. \cite{cd-11}, ending up with an operator that has the desired properties. In particular, it is possible to construct $\mathcal{T}_\mathcal{A}$ satisfying
\be\label{eq:norm_twist}
\la \mathcal{T}_\mathcal{A}\ra^{\otimes n} = \text{Tr}\l \rho^n_\mathcal{A}\r,
\ee
with $\la \dots \ra^{\otimes n}$ the replica state of $\mathcal{A}^{\otimes n}$ associated with the density matrix $\rho^{\otimes n}_{\mathcal{A}} \in \mathcal{A}^{\otimes n}$.

The uniqueness of the twist operator comes from the fact that $\mathcal{A}^{\otimes n}$ has a trivial center, since it is isomorphic to $\text{End}(\mathbb{C}^{dn})$, and we give an explanation below.
We first observe that, from Eq. \eqref{eq:Twist_op_def}, $\mathcal{T}_\mathcal{A}\mathcal{T}_\mathcal{A}^{\dagger}$ belongs to the center of $\mathcal{A}^{\otimes n}$; therefore, if $\mathcal{T}_\mathcal{A}\neq 0 $, it is in an invertible element of $\mathcal{A}^{\otimes n}$ and its inverse is proportional to $\mathcal{T}^\dagger_\mathcal{A}$. Let us now consider two non-zero twist operators $\mathcal{T}_\mathcal{A}, \hat{\mathcal{T}}_\mathcal{A}$ satisfying Eq. \eqref{eq:Twist_op_def}: $(\mathcal{T}_\mathcal{A})^{-1}\hat{\mathcal{T}}_{\mathcal{A}}$ has to be in the center of $\mathcal{A}^{\otimes n}$ and, since the latter is trivial, it implies $\mathcal{T}_\mathcal{A} \propto \hat{\mathcal{T}}_\mathcal{A}$. The only arbitrariness in the definition of $\mathcal{T}_\mathcal{A}$ is, therefore, the proportionality constant, which can be fixed unambiguously by requiring the normalisation condition \eqref{eq:norm_twist}.

At this point, we have an operator $\mathcal{T}_\mathcal{A}$ defined canonically at the algebraic level, without explicit reference to any local Hilbert space; in this way, we can give meaning to its dual under Kramers-Wannier duality, as done for the density matrix in the previous section. We start from the isomorphism in Eq. \eqref{eq:isom}, between two algebras. We extend it to an isomorphism of their replica as follows
\be
\begin{split}
\mathcal{A}^{\otimes n}&\rightarrow {\mathcal{A}'}^{\otimes n}\\
\mathcal{O} &\mapsto \mathcal{O}' \equiv T^{\otimes n}\mathcal{O}(T^\dagger)^{\otimes n}.
\end{split}
\ee
Following this procedure, we construct a dual twist operator associated with $\mathcal{A}'$ as
\be\label{eq:Renyi_twist}
\mathcal{T}_{\mathcal{A'}} \equiv T^{\otimes n}\mathcal{T}_{\mathcal{A}} \l T^{\dagger}\r^{\otimes n}.
\ee
The field above satisfies some natural expected properties. In particular, it allows representing the moments of the dual-density matrix associated with the dual algebra. A direct computation, based on Eqs. \eqref{eq:Renyi_twist} and \eqref{eq:equal_entropy}, gives indeed
\be
\la \mathcal{T}_\mathcal{A'} \ra^{' \otimes n} = \la T^{\otimes n}\mathcal{T}_\mathcal{A} (T^{\otimes n})^\dagger \ra^{' \otimes n} = \la \mathcal{T}_\mathcal{A}\ra^{\otimes n} = \text{Tr}\l \rho^n_{\mathcal{A}}\r= \text{Tr}\l \rho^n_{\mathcal{A}'}\r.
\ee

As an instructive example, we reinterpret the construction above in the context of the Ising Field Theory of Sec. \ref{sec:tfield_algebra}. The twist field $\mathcal{T}(x)$ in Eq. \eqref{eq:def_twist} is the twist operator associated with the algebra $\mathcal{A}$ generated by the insertion of $\varepsilon(y)$ and $\sigma(y)$ in the region $y>x$. In particular, it holds $\mathcal{T}(x)\sigma^{j}(y) = \sigma^{j+1}(y)\mathcal{T}(x)$ if $x<y$, holds. We can consider another algebra $\mathcal{A}'$, related to $\mathcal{A}$ via Kramers-Wannier duality, associated with the insertions of $\varepsilon(y)$ and $\mu(y)$ in $y>x$. We call $\mathcal{T}'(x)$ the associated (dual) twist operator, which satisfies
\be\label{eq:T1_comm_rel}
\mathcal{T}'(x)\mu^{j}(y) = \mu^{j}(y)\mathcal{T}'(x), \quad y<x.
\ee
In other words, $\mathcal{T}'$ \lq\lq sees\rq\rq $\mu$ as a local operator. $\mathcal{T}$ and $\mathcal{T}'$ are explicitly different, and we remark that the commutation relation in Eq. \eqref{eq:Tfield_mu} are distinct from those in Eq. \eqref{eq:T1_comm_rel}. In conclusion, there is no reason why the correlation functions of $\mathcal{T}$, associated with the usual notion of entanglement of the Ising model, should be equal in the ordered and disordered phases: we can only infer relations between $\mathcal{T}$ and $\mathcal{T}'$.

\section{Conclusions}
\label{Sec:conclusion}

In this work we studied the entropy of kink excitation, showing that universal results emerge in the limit of large regions: this finding extends previous studies on the entanglement content of quasiparticles. We provided a detailed analysis of specific states of a spin-1/2 chain and made use of the qubit-picture, where computations can be carried out explicitly with elementary methods. Furthermore, we discussed a field-theoretic framework and found that the discrepancy between particles/kinks is ultimately traced back to the algebra of twist fields and the corresponding local/semilocal operators. In particular, the disorder fields \lq\lq charge\rq\rq non-trivially the twist fields, as shown by Eq. \eqref{eq:Tfield_mu}, and a new family of twist fields (first introduced and characterised in Ref. \cite{cm-23}) is generated. Remarkably, while the qubit picture is simpler to deal with, the field-theoretic approach reveals the universal origin of the entropy difference between the kink states and the vacuum, which is far from obvious.

We pointed out the importance of the notion of \textit{entropy of algebras}, which is crucial to understanding the lack of correspondence of entanglement entropy under Kramers-Wannier duality. A related, albeit different, discussion found in the previous literature concerns the entropy of the Ising model compared to that of free Majorana fermions: while the two models are related (by Jordan Wigner transformation), a discrepancy in the two-interval entropy has been found in Refs. \cite{Fagotti-2010,Coser-2016}. The origin of the mismatch comes from the computation of the entropy of two distinct algebras: the one generated by local operators (say, Pauli matrices), and the one generated by fermions; this difference can be equivalently expressed in terms of spin-structure in the path integral formalism. We also mention recent results \cite{mf-23,mf-23a,fmz-24} on the presence of novel entanglement features in one-dimensional spin models mapped to free fermions, such as universal tripartite information and topological order, previously overlooked.

We remark that the main results of this work do not refer specifically to the Ising model and they do not make use of free fermionic techniques; the only important assumption is the spontaneous symmetry breaking of a $\mathbb{Z}_2$ symmetry with kink excitations interpolating between the vacua as a low-lying spectrum. In particular, our field-theoretic approach could be used to discriminate the entanglement content of deconfined kinks and their bound states appearing in the more exotic spectrum of the confining chain (Ising model with both transverse and longitudinal fields) of Ref. \cite{kt-24}. Furthermore, we expect generalisations to other systems with distinct symmetry-breaking patterns to be straightforward (e.g. the Potts model in its ordered phase): one should identify the algebraic relations between disordered fields and the twist fields and carry out the contractions to evaluate the correlation functions (as done in Sec. \ref{Sec:field-theoretic-approach}). In this respect, a promising framework to formulate a general theory, that incorporates both fields with disorder lines and duality lines attached to them (as for Kramers-Wannier) is that of \textit{Non-invertible generalised symmetries}, recently reviewed in Ref. \cite{Sakura-23}.

\medskip {\bf Acknowledgements:} 

LC acknowledges support from ERC  Starting grant 805252 LoCoMacro. LC thanks Vanja Maric and Michele Fossati for their insightful comments on the manuscript. MM is grateful for funding under the EPSRC Mathematical Sciences Doctoral Training Partnership EP/W524104/1. MM thanks Olalla A. Castro-Alvaredo, David Horvath and Fabio Sailis for the many useful discussions.

\begin{appendix}

\section{Details of some lattice and qubit calculations}
\label{app:lattice_qubit_calc}
In this appendix, we collect some technical calculations on the reduced density matrix of multi-kink states on the lattice and within the qubit picture.
\subsection{RDM of the lattice two-kink state in tripartite geometry}
\label{app:RDM of the lattice two-kink state}
Here, we derive the explicit expression for the reduced density matrix in Section \ref{sec:two-kink}, that is
\begin{align}
\rho^{(2)}_{A_1 \cup A_3} = \mathcal{N} \sum_{\substack{s_k=\pm \\ k=\ell_1+1,\dots,\ell_1+\ell_2}}\, \sum_{\substack{1 \le i_1 < j_1 \le L-1 \\ 1 \le i_2 < j_2 \le L-1}} &\prescript{}{A_2}{\braket{s_{\ell_1 +1},\dots,s_{\ell_1+\ell_2}|K_{+-}(i_1)K_{-+}(j_1)}}\nonumber \\
  \times &\braket{K_{+-}(i_2)K_{-+}(j_2)|s_{\ell_1 +1},\dots,s_{\ell_1+\ell_2}}_{A_2},
\end{align}
with $\mathcal{N} \equiv \frac{2}{(L-1)(L-2)}$ a normalisation constant. For convenience, we split the indices appearing in the sum into six subsets:
\begin{align*}
&\mathcal{S}_1 = \{(i,j)|i,j \in A_1, i < j \}, \nonumber \\
&\mathcal{S}_2 = \{(i,j)|i \in A_1, j \in A_2\setminus\{\ell_1 + \ell_2\}\}, \nonumber \\
&\mathcal{S}_3 = \{(i,j)|i \in A_1, j \in \{\ell_1 + \ell_2\} \cup A_3\setminus\{L\}\}, \nonumber \\
&\mathcal{S}_4 = \{(i,j)|i, j \in A_2\setminus\{\ell_1 + \ell_2\}, i < j \}, \nonumber \\
&\mathcal{S}_5 = \{(i,j)|i \in A_2\setminus\{\ell_1 + \ell_2\}, j \in \{\ell_1 + \ell_2\} \cup A_3\setminus\{L\} \}, \nonumber \\
&\mathcal{S}_6 = \{(i,j)|i, j \in \{\ell_1 + \ell_2\} \cup A_3\setminus\{L\}, i < j \}. \nonumber \\
\end{align*}
Then, we compute
\begin{align}
    &\prescript{}{A_2}{\braket{s_{\ell_1 +1},\dots,s_{\ell_1+\ell_2}|K_{+-}(i_1)K_{-+}(j_1)}} \nonumber \\
&= \l\delta_{(i_1, j_1) \in \mathcal{S}_1} \prod_{k \in A_2} \delta_{s_k,-}\r \ket{K_{+-}(i_1)K_{-+}(j_1)}_{A_1}\otimes \ket{+,\dots,+}_{A_3} \nonumber \\ &+ \l\delta_{(i_1, j_1) \in \mathcal{S}_2} \prod_{k = \ell_1+1}^{j_1} \delta_{s_k,-}\prod_{k = j_1+1}^{\ell_1+\ell_2} \delta_{s_k,+}\r \ket{K_{+-}(i_1)}_{A_1} \otimes \ket{+,\dots,+}_{A_3} \nonumber \\ &+\l\delta_{(i_1, j_1) \in \mathcal{S}_3}  \prod_{k \in A_2} \delta_{s_k,-}\r \ket{K_{+-}(i_1)}_{A_1}\otimes\ket{K_{-+}(j_1)}_{A_3}
\nonumber \\ &+\l\delta_{(i_1, j_1) \in \mathcal{S}_4}  \prod_{k = \ell_1+1}^{i_1} \delta_{s_k,+}\prod_{k = i_1+1}^{j_1} \delta_{s_k,-}\prod_{k = j_1+1}^{\ell_1 + \ell_2} \delta_{s_k,+}\r \ket{+,\dots,+}_{A_1}\otimes\ket{+,\dots,+}_{A_3}
\nonumber \\ &+\l\delta_{(i_1, j_1) \in \mathcal{S}_5}  \prod_{k = \ell_1+1}^{i_1} \delta_{s_k,+}\prod_{k = i_1+1}^{\ell_1 +\ell_2} \delta_{s_k,-}\r \ket{+,\dots,+}_{A_1}\otimes \ket{K_{-+}(j_1)}_{A_3}
\nonumber \\ &+\l\delta_{(i_1, j_1) \in \mathcal{S}_6}  \prod_{k \in A_2} \delta_{s_k,+}\r \ket{+,\dots,+}_{A_1}\otimes \ket{K_{+-}(i_1)K_{-+}(j_1)}_{A_3}.
\end{align}
In the above expression, to keep the formula as compact as possible we have made the identifications $\ket{K_{+-}(\ell_1)}_{A_1}\equiv \ket{+,\dots,+}_{A_1}$, $\ket{K_{-+}(\ell_1+\ell_2)}_{A_3}\equiv \ket{+,\dots,+}_{A_3}$, $\ket{K_{+-}(i)K_{-+}(\ell_1)}_{A_1}\equiv \ket{K_{+-}(i)}_{A_1}$, and  $\ket{K_{+-}(\ell_1+\ell_2)K_{-+}(i)}_{A_3}\equiv \ket{K_{-+}(i)}_{A_3}$. The RDM is given by the sum of eight blocks, obtained by matching the delta constraints coming from the two matrix elements:
\begin{align}
\rho^{(2)}_{A_1 \cup A_3} &= \mathcal{N}\sum_{\substack{(i_1,j_1) \in \mathcal{S}_1\\ (i_2,j_2) \in \mathcal{S}_1}} \ket{K_{+-}(i_1)K_{-+}(j_1)}_{A_1} \prescript{}{A_1}{\bra{K_{+-}(i_2)K_{-+}(j_2)}} \otimes \ket{+\dots,+}_{A_3} \prescript{}{A_3}{\bra{+,\dots,+}} \nonumber \\ &+ \mathcal{N}\sum_{\substack{(i_1,j_1) \in \mathcal{S}_6\\ (i_2,j_2) \in \mathcal{S}_6}} \ket{+\dots,+}_{A_1} \prescript{}{A_1}{\bra{+,\dots,+}} \otimes \ket{K_{+-}(i_1)K_{-+}(j_1)}_{A_3} \prescript{}{A_3}{\bra{K_{+-}(i_2)K_{-+}(j_2)}}  \nonumber \\ 
&+ \mathcal{N}\sum_{\substack{(i_1,j_1) \in \mathcal{S}_1\\ (i_2,j_2) \in \mathcal{S}_6}}  \ket{K_{+-}(i_1)K_{-+}(j_1)}_{A_1} \prescript{}{A_1}{\bra{+,\dots,+}} \otimes \ket{+\dots,+}_{A_3}\prescript{}{A_3}{\bra{K_{+-}(i_2)K_{-+}(j_2)}}  \nonumber \\ 
&+ \mathcal{N}\sum_{\substack{(i_1,j_1) \in \mathcal{S}_6\\ (i_2,j_2) \in \mathcal{S}_1}} \ket{+\dots,+}_{A_1} \prescript{}{A_1}{\bra{K_{+-}(i_2)K_{-+}(j_2)}} \otimes \ket{K_{+-}(i_1)K_{-+}(j_1)}_{A_3} \prescript{}{A_3}{\bra{+,\dots,+}}  \nonumber \\ 
&+ \mathcal{N}\sum_{\substack{(i_1,j_1) \in \mathcal{S}_3\\ (i_2,j_2) \in \mathcal{S}_3}} \ket{K_{+-}(i_1)}_{A_1} \prescript{}{A_1}{\bra{K_{+-}(i_2)}} \otimes \ket{K_{-+}(j_1)}_{A_3} \prescript{}{A_3}{\bra{K_{-+}(j_2)}}  \nonumber \\ 
&+ \mathcal{N}(\ell_2 - 1)\sum_{i_1, i_2 \in A_1} \ket{K_{+-}(i_1)}_{A_1} \prescript{}{A_1}{\bra{K_{+-}(i_2)}} \otimes  \ket{+\dots,+}_{A_3} \prescript{}{A_3}{\bra{+,\dots,+}}  \nonumber \\ 
&+ \mathcal{N}(\ell_2 - 1)\sum_{j_1, j_2 \in \{\ell_1 + \ell_2\}\cup A_3 \setminus L}  \ket{+\dots,+}_{A_1} \prescript{}{A_1}{\bra{+,\dots,+}} \otimes  \ket{K_{-+}(j_1)}_{A_3} \prescript{}{A_3}{\bra{K_{-+}(j_2)}}  \nonumber \\
&+ \mathcal{N}\frac{(\ell_1 - 1)(\ell_2 -2)}{2}\ket{+\dots,+}_{A_1} \prescript{}{A_1}{\bra{+,\dots,+}} \otimes \ket{+\dots,+}_{A_3} \prescript{}{A_3}{\bra{+,\dots,+}}.
\end{align}

\subsection{Qubit description of three- and four-kink states}
\label{app:qubit_34_kinks}

Here we diagonalise the reduced density matrix of the states with $N=3,4$ kinks in Eq. \eqref{eq:N_kink_state_qubit}. These states are
\begin{align}
\label{3K_state}
\ket{\Psi^{(3)}} &= \sqrt{6 r_1 r_2 r_3} \ket{1^+, 1^-, 1^+} \nonumber \\
&+ \sqrt{3 r_1^2 r_2}\ket{2^+,1^+,-} + \sqrt{3 r_1 r_2^2}\ket{1^+,2^-,-} \nonumber \\
&+ \sqrt{3 r_1^2 r_3}\ket{2^+,+,1^+} + \sqrt{3 r_1 r_3^2}\ket{1^+,-,2^-} \nonumber \\
&+ \sqrt{3 r_2^2 r_3}\ket{+,2^+,1^+} + \sqrt{3 r_2 r_3^2}\ket{+,1^+,2^-} \nonumber \\
&+ r_1^{3/2} \ket{3^+,-,-} + r_2^{3/2} \ket{+,3^+,-} + r_3^{3/2} \ket{+,+,3^+},
\end{align}
\begin{align}
\label{4K_state}
\ket{\Psi^{(4)}} &= \sqrt{12 r_1^2 r_2 r_3} \ket{2^+, 1^+, 1^-} + \sqrt{12 r_1 r_2^2 r_3} \ket{1^+, 2^-, 1^-} + \sqrt{12 r_1 r_2 r_3^2} \ket{1^+, 1^-, 2^+}\nonumber \\
&+ \sqrt{4 r_1 r_2^3}\ket{1^+,3^-,+} + \sqrt{6 r_1^2 r_2^2}\ket{2^+,2^+,+} + \sqrt{4 r_1^3 r_2}\ket{3^+,1^-,+}\nonumber \\
&+ \sqrt{4 r_1 r_3^3}\ket{1^+,-,3^-} + \sqrt{6 r_1^2 r_3^2}\ket{2^+,+,2^+} + \sqrt{4 r_1^3 r_3}\ket{3^+,-,1^-}\nonumber \\
&+ \sqrt{4 r_2 r_3^3}\ket{+,1^+,3^-} + \sqrt{6 r_2^2 r_3^2}\ket{+,2^+,2^+} + \sqrt{4 r_2^3 r_3}\ket{+,3^+,1^-}\nonumber \\
&+ r_1^2 \ket{4^+,+,+} + r_2^2 \ket{+,4^+,+} + r_3^2 \ket{+,+,4^+}.
\end{align}
We trace out the degrees of freedom of $A_2$, obtaining the diagonal from of the two RDMs
\begin{align}
\label{diagona_RDM_3K}
\rho_A^{(3)}&= r_2^3 \ket{\Phi_0}\bra{\Phi_0} + 3r_1 r_2^2 \ket{\Phi_1^-}\bra{\Phi_1^-} + 3r_3 r_2^2 \ket{\Phi_1^+}\bra{\Phi_1^+} \nonumber \\&+ 6 r_1 r_2 r_3 \ket{\Phi_2^-}\bra{\Phi_2^-} + 3r_2(r_1^2 + r_3^2) \ket{\Phi_2^+}\bra{\Phi_2^+} \nonumber \\
&+ r_1 (r_1^2 + 3r_3^2) \ket{\Phi_3^-}\bra{\Phi_3^-} +  r_3(3r_1^2 + r_3^2) \ket{\Phi_3^+}\bra{\Phi_3^+}, 
\end{align}
and
\begin{align}
\label{diagona_RDM_4K}
\rho_A^{(4)}&= r_2^4 \ket{\Phi_0}\bra{\Phi_0} + 4r_1 r_2^3 \ket{\Phi_1^-}\bra{\Phi_1^-} + 4r_3 r_2^3 \ket{\Phi_1^+}\bra{\Phi_1^+} \nonumber \\&+ 12 r_1 r_2^2 r_3 \ket{\Phi_2^-}\bra{\Phi_2^-} + 6r_2^2(r_1^2 + r_3^2) \ket{\Phi_2^+}\bra{\Phi_2^+} \nonumber \\
&+ 4 r_1 r_2(r_1^2 + 3r_3^2) \ket{\Phi_3^-}\bra{\Phi_3^-} + 4 r_2 r_3(3r_1^2 + r_3^2) \ket{\Phi_3^+}\bra{\Phi_3^+} \nonumber \\ &+ 4 r_1 r_3(r_1^2 + r_3^2) \ket{\Phi_4^-}\bra{\Phi_4^-} + (r_1^4 + r_3^4 + 6r_1^2 r_3^2) \ket{\Phi_4^+}\bra{\Phi_4^+}.
\end{align}
The eigenvectors of $\rho_A^{(3)}$ are:
\begin{itemize}
    \item $\ket{\Phi_0} = \ket{+,-}$ (no kinks in $A$)
    \item $\ket{\Phi_1^-} = \ket{1^+,-}$ (one kink in $A_1$, no kinks in $A_3$)
    \item $\ket{\Phi_1^+} = \ket{+,1^+}$ (no kinks in $A_1$, one in $A_3$)
    \item $\ket{\Phi_2^-} = \ket{1^+,1^+}$ (one kink in $A_1$, one in $A_3$)
    \item $\ket{\Phi_2^+} = \frac{r_1 \ket{2^+,-}+r_3\ket{+,2^-}}{\sqrt{r_1^2 +r_3^2}}$ (one magnon in $A_1$ or one magnon in $A_3$)
    \item $\ket{\Phi_3^-}= \frac{r_1 \ket{3^+,-}+\sqrt{3}r_3\ket{1+,2^-}}{\sqrt{r_1^2 +3r_3^2}}$ (3 kinks in $A_1$ or one kink in $A_1$ and one magnon in $A_3$)
    \item $\ket{\Phi_3^+}= \frac{\sqrt{3}r_1 \ket{2^+,1^+}+r_3\ket{+,3^+}}{\sqrt{3r_1^2 +r_3^2}}$ (1 magnon in $A_1$ and one kink in $A_3$ or 3 kinks in $A_3$),
\end{itemize}
while those of $\rho_A^{(4)}$ are:
\begin{itemize}
    \item $\ket{\Phi_0} = \ket{+,+}$ (no kinks in $A$)
    \item $\ket{\Phi_1^-} = \ket{1^+,+}$ (one kink in $A_1$)
    \item $\ket{\Phi_1^+} = \ket{+,1^-}$ (no kinks in $A_1$, one in $A_3$)
    \item $\ket{\Phi_2^-} = \ket{1^+,1^-}$ (one kink in $A_1$, one in $A_3$)
    \item $\ket{\Phi_2^+} = \frac{r_1 \ket{2^+,+}+r_3\ket{+,2^+}}{\sqrt{r_1^2 +r_3^2}}$ (one magnon in $A_1$ or one magnon in $A_3$)
    \item $\ket{\Phi_3^-}= \frac{r_1 \ket{3^+,+}+\sqrt{3}r_3\ket{1+,2^+}}{\sqrt{r_1^2 +3r_3^2}}$ (3 kinks in $A_1$ or one kink in $A_1$ and one magnon in $A_3$)
    \item $\ket{\Phi_3^+}= \frac{\sqrt{3}r_1 \ket{2^+,1^+}+r_3\ket{+,3^+}}{\sqrt{3r_1^2 +r_3^2}}$ (1 magnon in $A_1$ and one kink in $A_3$ or 3 kinks in $A_3$)
    \item $\ket{\Phi_4^-} = \frac{r_1 \ket{3^+,1^-}+r_3\ket{1^+,3^-}}{\sqrt{r_1^2 +r_3^2}}$ (3 kinks in $A_1$ and one in $A_3$ or 1 kink in $A_1$ and three in $A_3$ )
    \item $\ket{\Phi_4^+}=\frac{r_1^2 \ket{4^+,+}+\sqrt{6r_1^2 r_3^2}\ket{2^+,2^+}+ r_3^2\ket{+,4^+}}{\sqrt{r_1^4 +r_3^4 + 6r_1^2 r_3^2}}$ (2 magnons in $A_1$ or 2 magnons in $A_3$ or 1 magnon in $A_1$ and one magnon in $A_3$).
\end{itemize}
The expressions of the eigenstates show how the superposition mechanism explained at the end of Section \ref{sec:lattice_qubit} works in detail. It is easy to check that the eigenvalues in Eqs. \eqref{diagona_RDM_3K} and \eqref{diagona_RDM_4K} match the formulae \eqref{eq:0_kinks_in_A_eigenvalue}, \eqref{eq:N_A_kinks_in_A_eigenvalue}.

\section{More on $C^{*}$-algebras and twist operators}\label{app:cstar}

In this appendix, we first review some elementary properties of the finite-dimensional $C^{*}$-algebras, and then we discuss some additional details regarding the notions of twist operators and entropy. We remark that infinite-dimensional algebras arise systematically in rigorous treatments of the observables of infinite dimensional spin lattices \cite{br-87} and QFT \cite{Witten-18}, but their treatment is beyond the purpose of this work. However, in this context, we consider the algebra of finite regions to be finite-dimensional, potentially incorporating appropriate ultraviolet regularisation when dealing with quantum field theories.

The first important result is a classification theorem for those algebras, which goes under the name of \textit{Wedderburn–Artin theorem} (see for instance \cite{lam1991first,Nicholson-93}). It states that every finite-dimensional $C^{*}$-algebra $\mathcal{A}$ is isomorphic to a direct sum of full matrix algebras, namely
\be\label{eq:decomposition}
\mathcal{A} \simeq \bigoplus_{\lambda} \text{End}\l \mathbb{C}^{d_\lambda}\r.
\ee
Each term in the previous sum is called a \textit{factor}, and it is (isomorphic to) an algebra of $d_\lambda\times d_\lambda$ complex matrices. The \textit{center} of the algebra $\mathcal{A}$, defined as
\be
\mathcal{Z}(\mathcal{A}) \equiv \left \{ c \in \mathcal{A}  |\, [c,a] = 0 \quad  \forall a \in \mathcal{A}\right \},
\ee
is generated by the identity operators of each factor
\be
\mathcal{Z}(\mathcal{A}) = \text{Span}\left\{ 1_\lambda  \in \text{End}\l \mathbb{C}^{d_\lambda}\r \right\}.
\ee
In the case of a single factor, which is the usual scenario for the algebra of regions of spin chains, the center is trivial as it is generated by the identity matrix. Nonetheless, in other contexts, as for lattice gauge theories \cite{chr-13}, the physical observables may display multiple factors, and we refrain from making specific assumptions in this regard.

An important concept in defining the density matrix and the entropy is the (intrinsic) notion of trace. In the context of algebras, a trace is defined as a positive linear functional
\be\label{eq:trace}
\begin{split}
\mathcal{A} &\rightarrow \mathbb{C},\\
a &\rightarrow \text{Tr}\l a\r,
\end{split}
\ee
which satisfies $\text{Tr}(ab-ba) = 0$. Without further hypothesis, it might not be obvious whether a trace exists or it is unique (up to a proportionality constant). The first important result is that, whenever $\mathcal{A}$ is a \textit{full matrix algebra}, say
\be
\mathcal{A} = \text{End}\l \mathbb{C}^d\r,
\ee
then the trace is uniquely defined up to a proportionality constant, and it corresponds to the usual notion of the trace of ($d\times d$ complex) matrices. To prove the statement, it is sufficient to observe that the image of the commutator
\be
[a,b]\equiv ab-ba,
\ee
has dimension $d^2-1$ \footnote{
Given an orthonormal basis $\{\ket{i}\}$ one has $[\ket{i}\bra{j},\ket{j}\bra{k}] = \ket{i}\bra{k}$ for $i \neq k$, and $[\ket{i}\bra{j},\ket{j}\bra{i}] = \ket{i}\bra{i}-\ket{j}\bra{j}$. The linear combinations of those matrices which are obtained via commutators generate the space of traceless matrices.
} and it is in direct sum with the span of the identity matrix. Therefore, to define a trace satisfying Eq. \eqref{eq:trace}, it is sufficient to specify its value for the identity operator. The most common normalisation, associated with the usual trace of matrices, is
\be
\text{Tr}\l 1\r=d.
\ee
For algebras with multiple factors, many traces can be constructed in principle, and they are identified by the values of the elements in the center; a natural choice is nonetheless
\be
\text{Tr}\l 1_\lambda \r=d_\lambda.
\ee

It is worth commenting on the entropy for algebras with multiple factors (see details in Ref. \cite{ch-22}). Given a state $\la \dots\ra$ for $\mathcal{A}$ in \eqref{eq:decomposition}, we split it as
\be
\la \dots \ra = \sum_\lambda p_\lambda \la \dots \ra_{\lambda}.
\ee
Here $p_\lambda \equiv \frac{1}{\la 1_\lambda\ra}$ is the probability of the factor $\lambda$; $\la \dots\ra_\lambda$ is the state conditioned to the factor $\lambda$, which satisfies $\la 1_\lambda\ra_\lambda =1$ and $\la \mathcal{O}\ra_\lambda =0$ if $\mathcal{O} \in \mathcal{A}_{\lambda'}$ ($\lambda'\neq \lambda$). Similarly, the density matrix associated with $\la \dots\ra$ is decomposed as
\be
\rho_{\mathcal{A}} = \sum_\lambda p_\lambda \rho_{\mathcal{A}_\lambda},
\ee
with $\rho_{\mathcal{A}_\lambda}$ the density matrix of $\la \dots\ra_\lambda$. A simple calculation \cite{ch-22} gives the entropy of the state as
\be
S(\rho_\mathcal{A}) = \sum_\lambda p_\lambda S(\rho_{\mathcal{A}_\lambda}) - \sum_\lambda p_\lambda \log p_\lambda.
\ee
We mention that a similar decomposition for the entropy arises whenever the density matrix displays a block-diagonal structure, as pointed out in Refs. \cite{Greiner-19,xas-18} for $U(1)$ conserving systems. The computation of the moments of the reduced density matrix, and thus of the R\'enyi entropies, can be given in terms of twist operator for algebras with non-trivial center as well, as we explain below.

First, one defines $\mathcal{T}_{\mathcal{A}_\lambda}$ as twist operators associated with $\mathcal{A}_\lambda$ and belonging to $\mathcal{A}^{\otimes n}_\lambda$: no issues arise here since the center of $\mathcal{A}_\lambda$ is trivial. Then, one introduces
\be
\mathcal{T}_{\mathcal{A}} \equiv \sum_\lambda \mathcal{T}_{\mathcal{A}_\lambda} \in \mathcal{A}^{\otimes n},
\ee
which acts as a replica cyclic permutation and it does not connect distinct factors. We define the replica state $\la \dots\ra^{\otimes n}$ as the state of $\mathcal{A}^{\otimes n}$ associated with the density matrix $\rho^{\otimes n}_{\mathcal{A}} \in \mathcal{A}^{\otimes n}$. A direct computation gives
\be
\begin{split}
\text{Tr}\l \mathcal{T}_\mathcal{A}\rho^{\otimes n}_{\mathcal{A}}\r = &\sum_{\lambda,\lambda_1,\dots,\lambda_n} p_{\lambda_1}\dots p_{\lambda_n}
\text{Tr}\l \mathcal{T}_{\mathcal{A},\lambda} \l\rho_{A,\lambda_1}\otimes \dots \otimes \rho_{A,\lambda_n}\r\r = \sum_{\lambda} p^n_\lambda \text{Tr}\l \mathcal{T}_{\mathcal{A}_\lambda} \rho^{\otimes n}_{\mathcal{A},\lambda}\r \\=
&\sum_{\lambda} p^n_\lambda \text{Tr}\l \rho^n_{\mathcal{A},\lambda}\r = \text{Tr}\l \rho^n_{\mathcal{A}}\r,
\end{split}
\ee 
where we used the fact that contributions associated to distinct factors vanish and only the terms with $\lambda=\lambda_1=\dots = \lambda_n$ remain.

Finally, we give some remarks regarding the existence of twist operators and their generalisation. In the main text, we mention that a microscopic construction for an operator that implements the replica shift $j\rightarrow j+1$ via Eq. \eqref{eq:Twist_op_def} can be provided explicitly. 
In different contexts, other twist operators are found to be useful in the study of entanglement, as for example composite fields with $\mathbb{Z}_2$ lines attached to them, as for the fields in Sec. \ref{sec:tfield_algebra} and those of Ref. \cite{hc-20}. Therefore, a natural question to pose is whether a given transformation of observables can always be implemented via a certain operator. More precisely, given
\be
\phi: \mathcal{A} \rightarrow \mathcal{A},
\ee
an algebra automorphism (that is $\phi(ab) = \phi(a)\phi(b)$, $ \phi(a^\dagger) = \phi(a)^\dagger$) we look for an element $\mathcal{T}^\phi \in \mathcal{A}$ satisfying
\be\label{eq:tphi}
\mathcal{T}^\phi a = \phi(a) \mathcal{T}^\phi, \quad \forall a \in \mathcal{A}.
\ee
The cyclic permutation of the replica model in Eq. \eqref{eq:Twist_op_def} is indeed a specific example of this general problem. In the case of algebras with trivial central, a general result, under the name of \textit{Skolem–Noether theorem} \cite{Skolem-27}, ensures that such $\mathcal{T}^\phi$ always exist: in other words, every automorphism can be realised as an inner automorphism via $\mathcal{T}^\phi a (\mathcal{T}^{\phi})^{-1} = \phi(a)$. This result is far from being trivial, and we are not aware of any simple constructive proof. Nonetheless, a proof of the unicity (up to constants) is definitely simpler, as it follows the same argument of Sec. \ref{sec:d_tfield}, and it relies on the triviality of the center of $\mathcal{A}$. Lastly, we point out that, in the presence of a non-trivial center, not only the unicity but also the existence of an operator $\mathcal{T}^\phi$ satisfying \eqref{eq:tphi} is not guaranteed a priori. A simple example is provided by the commutative algebra generated by two central elements $a,b$ (that is, $\mathcal{A}\simeq \text{End}(\mathbb{C})\oplus \text{End}(\mathbb{C})$) satisfying $a^2=a,b^2=b$: the map $\phi(a) = b, \phi(b) = a$ is an automorphism, but it clearly cannot be realised as an inner automorphism since the algebra is commutative. 

\end{appendix}

\printbibliography

\end{document}